\documentclass{article}
\font\cero=cmss10 scaled 1728 
\usepackage{tensor}
\usepackage{verbatim}
\usepackage{amsfonts}
\usepackage{graphicx}
\usepackage{amssymb}
\usepackage{amsmath}
\usepackage{amssymb}
\usepackage{braket}
\usepackage{comment}
\usepackage{float}
\usepackage{cite}
\usepackage{notoccite}
\usepackage[utf8]{inputenc}
\usepackage{hyperref}

\usepackage[sort&compress,square,numbers]{natbib}

\begin{document}
\begin{flushleft}
{\cero Asymptotic entangled states from the dissipative interaction of two charged fields}\\
\end{flushleft} 
{\sf R. Cartas-Fuentevilla, O. Cruz-Limón* and C. Ramírez-Romero*}\\
{\it Instituto de F\'{\i}sica, Universidad Aut\'onoma de Puebla,
Apartado postal J-48 72570 Puebla Pue., M\'exico.} \\
{*\it Facultad de Ciencias Físico Matemáticas, Benemérita Universidad Autónoma de Puebla. P.O. Box 165, 72570 Puebla, México.}\\ 

\noindent
ABSTRACT: We construct a field theory for dissipative systems using a hypercomplex ring formalism that reproduces naturally the effective doubling field formulation of dissipative systems of thermal field theory. The system is quantized by a noncanonical ansatz that gives the unitarity of the system is conserved. Asymptotically entangled states are constructed to introduce the study of the ergodicity of systems that undergo entanglement.\\


\section{Introduction}

Quantum mechanics has the very particular strange feature that is also surprising, entanglement. This phenomenon is present in various areas as quantum optics \cite{1.1.1}, quantum field theory (QFT) \cite{1.1.2}, AdS/CFT correspondence \cite{33}, and it is an essential part for the development of quantum information theory, and for the emergence of new technologies related to it. For instance, in \cite{1.1} the authors develop new techniques to entangle disparate electromagnetic fields that range from microwave radiation to optical beams; if this entanglement was achieved, it would be closer to solving the problem that arises when entanglement is shared between two separate quantum computers. Another method that has been used to generate entanglement  is to use the non-local Seebeck effect \cite{1.2}, which consists of generating a thermoelectric current through a temperature difference by using a quantum splitter, and it is found that certain processes such as Copper pair splitting (CPS) and elastic co-tunneling (EC) can contribute to this current, achieving a tuning between CPS and EC, which allows us testing fundamental theories related to entanglement and heat transport in graphene systems.
On the other hand, in addition of being able to generate the entanglement it is important to measure it; for example in \cite{1.3}, the authors propose a test for entanglement in scattering of chemical reactions.\\

\noindent
The entanglement has been generated artificially, but in most cases it is generated naturally, often being an unwanted effect. A natural way in which entanglement is generated is as an effect of dissipation, which causes that populations of the quantum states are modified due to the interaction with the environment, which in turn causes a phase shift \cite{1.9}; these two actions correspond to the result of the intertwining between the degrees of freedom of the environment and the system of interest; in a system with these characteristics, the unitarity of the dynamics is lost \cite{1.10}. On the other hand, any small alteration that interferes with the unitarity of the quantum evolution is undesirable, since quantum states lose their coherence, that is, quantum decoherence arises. Decoherence is an effect that cannot be avoided, it arises when a quantum system is coupled to its environment; this phenomenon is often undesirable, since it causes the information of the system of interest to be lost. But not everything is bad with decoherence, there is also some interest in this phenomenon, as discussed in \cite{1.14}.\\

\noindent
For a long time, it has been of interest to study the quantum dynamics of a particle that is coupled with its environment, and several methods have been developed. A well-known and widely used method is the doubling of degrees of freedom; in \cite{1.8} the authors make use of this duplication to review the general features of a dissipative quantum model of the brain and discuss how QFT phase correlations and entanglement are achieved for modeling functional brain activity. On the other hand, for systems coupled to an environment, the thermodynamic descriptions assume that any system of any finite size, being in contact with a thermal reservoir with a temperature \(T\), will reach a state of thermal equilibrium after a period of time, which will maintain the same temperature \(T\). When there are dissipative processes that occur at very long times, it is normal to think that this thermalization occurs; however, this is not always the case. On the one hand, it is well known that in coherent quantum mechanics, there are time-periodic modulations that can lead to non-equilibrium asymptotic states, which are known as Floquet states. On the other hand, there are cycle-stationary states, periodic states in time; the latter introduce the ergodic theory to study dissipative systems \cite{1.8.9}.\\
\noindent
This work is organized as follows: in section \ref{section 2} we describe the formalism of the ring of hypercomplex numbers, essential objects in the construction of our dissipative model. In section \ref{sec3} we start by generalizing a Lagrangian consistent with the thermal field dynamics formalism. This generalization allows us to incorporate charge to the hyperbolic field theory realized in \cite{25}. In section \ref{sec4} we give a quantization ansatz that is not a canonical one, and in turn, the new commutation relations obtained will allow to eliminate dissipative factors that are not desired in dissipative theories \cite{39} \cite{39.1}.
In section \ref{sec5} we construct the Hamiltonian operator and we consider specific geometries for the total system as examples; we also consider the charge operator, from which a new hypercomplex description for the Noether charge will emerge due to the dissipation. Finally, we construct the evolution operator, giving way to the construction of the Matsubara formalism.
For section \ref{section 6} a definition of the vacuum through certain idempotent basis in the hyperbolic ring is considered and the expected values for quantum observables are calculated.
In section \ref{section 7}, the asymptotic entangled states are constructed considering that the geometry of the total system can be finite or infinite, which will lead us to consider the ergodic theory. Finally, in section \ref{conclusions} we present the conclusions.\\

\noindent
This work is based on the reference \cite{25}, in which the authors consider the Lagrangian formulation used in \cite{33} but within the formalism of a pure hyperbolic ring. They present new commutation relations that reveal an underlying noncommutative field theory that arises naturally due to the algebraic structure of the hyperbolic ring; in turn, the hyperbolic rotations reveal an underlying internal symmetry for the dissipative dynamics. The development of the work at hand is through the steps of \cite{25}, but now implementing an extension of the formalism, namely, the complex elliptic unit \(i\) is incorporated to the purely hyperbolic unit used in that work, obtaining thus a hypercomplex formalism. Contributions of the formalism at hand have opened the door to delve into other topics such as: the realization of holography through a grand partition function with duplication of fields, the relationship between the geometry of a dissipative system and ergodicity, the entanglement entropy and contributions of corrections to the Noether charge; these topics will be considered in future works.

\section{The hypercomplex numbers \(\mathbb{H}\)}
\label{section 2}
We begin by describing the formalism of hypercomplex numbers. In the formalism used in \cite{25} the pure hyperbolic numbers have the form,
\begin{eqnarray}
    \mathbb{P} \equiv \left\{w=x+jy; \quad j^2=1, \quad \bar{j}=-j; \quad x,y \in \mathbb{R} \right\}, 
\label{1}
\end{eqnarray}
which can be generalized by promoting the real numbers in (\ref{1}) to ordinary complex numbers, that is, \(x \rightarrow x+iy\) and \(y \rightarrow u+iv\), thus
\begin{eqnarray}
\begin{aligned}
\xi&=x+ iy + ju + ijv,           &  \bar{\xi}&=x - iy - ju + ijv;              &  x,y,u,v& \in \mathbb{R}.
\end{aligned}
\label{2}
\end{eqnarray}
We denote the ring of these numbers of four real components as \(\mathbb{H}\); this will lead, at a quantum level, to the existence of four bosons, whose excited states will allow us to describe the entangled states between the system of interest and the environment. This new extended ring has the following properties for the complex units; for the hyperbolic unit we have \(j^{2}=1\) and \(\overline{j}=-j\), for the standard complex unit \(i^{2}=-1\) and \(\overline{i}=-i\), additionally we have a new complex unit, composed of a hybrid term \(ij\), with the properties \((ij)^{2}=-1\), \(ij=ji\) and \(\overline{ij}=ij\). The ring \(\mathbb{H}\) has as subsets the pure hyperbolic numbers \(\mathbb{P}\) used in \cite{25} and the usual complex numbers \(\mathbb{C}\). The modulus of a hypercomplex number is given by \cite{24},
\begin{eqnarray}
    \xi \bar{\xi} = x^2 + y^2 - u^2 - v^2 + 2ij(xv-yu).  
\label{3}
\end{eqnarray}
This expression is invariant under the group \(U(1)\) by the action $\xi\rightarrow\xi e^{i\theta}$ and, under the group \(SO(1,1)\), which corresponds to hyperbolic rotations $\xi\rightarrow\xi e^{j\chi}$; hence (\ref{3}) is invariant under \(\xi \rightarrow \xi \, e^{j \chi} e^{i \theta}\), where the bi-complex phase can be expressed as,
\begin{eqnarray}
\begin{aligned}
e^{i \alpha} e^{j \beta} \equiv e^{i \alpha + j \beta} = \cos{(\alpha)} \cosh(\beta) + i\, \sin(\alpha) \cosh(\beta) + j \, \cos(\alpha) \sinh(\beta) + ij \, \sin(\alpha) \sinh(\beta).
\label{4}
\end{aligned}
\end{eqnarray}
Furthermore, the idempotent elements corresponding to the usual complex numbers are well know, namely, \(0\) and \(1\); the additional idempotent elements of \(\mathbb{H}\) are,
\begin{eqnarray}
\begin{aligned}
    J^{+}&= \frac{1}{2}(1+j), \quad (J^{+})^{n}=J^{+};\\
    J^{-}&= \frac{1}{2}(1-j), \quad (J^{-})^{n}=J^{-}, \quad n=1,2,3,...; 
\label{5}
\end{aligned}
\end{eqnarray}
and they will be constantly used; these elements work as orthogonal projectors; in addition, they eliminate to each other and are complex conjugates,
\begin{eqnarray}
\begin{aligned}
J^{+} J^{-}&=0,   \hspace{1cm}   &   (J^{+})^{\ast}&=J^{-}. 
\label{6}
\end{aligned}
\end{eqnarray}
Other properties of hyperbolic numbers that will be used in this work are detailed in section 2 in \cite{25}.


\section{The dissipative system and its Lagrangian formalism}
\label{sec3}
The basis of this work is the formulation of thermal fields that doubles the number of degrees of freedom \cite{1.17}, in turn, this formulation is used in \cite{25}, where the total system is represented by a real field \(\Phi\) for the system of interest \((A)\), and a second real field \(\Psi\) representing the thermal bath or environment \((B)\); here we will continue with that representation for the subsystems. In this article we generalize the formulation in \cite{25} by promoting the \(\Phi\) and \(\Psi\) fields to charged fields, according to the generalization described in Eq.(\ref{1}). The starting point is the following Lagrangian that includes a mass term, which will undergo also dissipation that not was considered in \cite{25}. The mass term has in general the form \(m_{1}^{2} + ijm_{2}^{2}\), but the corresponding dispersion relation will force \(m_{2}=0\), see Eq.(\ref{17}),
\begin{eqnarray}
    \mathcal{L}(\Omega, \overline{\Omega})=\frac{1}{2}\int dx^{d} \left [ \partial_{\mu}\Omega \cdot \partial^{\mu}\overline{\Omega} + \frac{\gamma}{2} (j\Omega \dot{\overline{\Omega}} + c.c)- m^2\Omega \overline{\Omega} \right ],  
\label{7}
\end{eqnarray}
where the field \(\Omega\) was a pure hyperbolic field of the form \(\Omega=\Phi + j \Psi\), but in the case at hand, the field \(\Omega\) will be generalized by a hypercomplex field, that is: \(\Phi \rightarrow \phi_{1}+i\phi_{2}\) and \(\Psi \rightarrow \psi_{1}+i\psi_{2}\), then \(\Omega\) takes the form,
\begin{eqnarray}
\begin{aligned}
\Omega&=\phi_{1} + i \phi_{2} + j\psi_{1} + ij \psi_{2};  &  \hspace{1.5cm} \phi_{1},\phi_{2},\psi_{1},\psi_{2} \in \mathbb{R}.
\label{8}
\end{aligned}
\end{eqnarray}
Since the \(\Omega\) field has been generalized, the additional symmetry \(U(1) \times SO(1,1)\) arises. Then the elements of the Lagrangian (\ref{7}) have the form (\ref{3}), 
\begin{eqnarray}
\begin{aligned}
\partial_\mu \Omega \cdot \partial^\mu \overline{\Omega} &= (\partial_\mu \, \phi_1)^2 + (\partial_\mu \, \phi_2)^2 - (\partial_\mu \, \psi_1)^{2} - (\partial_\mu \, \psi_2)^2
+ 2ij \, (\partial_\mu \, \phi_1\, \partial_\mu \, \psi_2 - \partial_\mu \, \phi_2 \, \partial_\mu \, \psi_1),
\label{9}
\end{aligned}
\end{eqnarray}
similarly for the terms \(\gamma(j\Omega \dot{\overline{\Omega}}+ c.c)\) and (\(m^2 \,\Omega \overline{\Omega}\)). The hybrid term of the form \(ij\) in (\ref{9}) can be understood as an interacting term between the two charges fields \citep{18}. On the other hand, using the idempotent bases (\ref{5}), the field (\ref{8}) can be expressed as, 
\begin{eqnarray}
    \Omega=J^{+} \Omega^{+} +J^{-}\Omega^{-},
\label{10}
\end{eqnarray}
where,
\begin{eqnarray}
\begin{aligned}
\Omega^{+}&=(\Phi+\Psi),   &  \Omega^{-}&= (\Phi-\Psi); & \hspace{0.3cm} \Phi=\phi_{1} + i \phi_{2}, \quad \Psi=\psi_{1} + i \psi_{2},
\label{11}
\end{aligned}
\end{eqnarray}
which are standard complex fields.
\noindent
When the equations of motion are calculated in the basis \((J^{+}, J^{-})\), the correspondence of each equation of motion in such a basis is notorious, namely, one for the system of interest \((J^{+})\) and, the other for the environment \((J^{-})\); this is due to the annihilation property (\ref{6}). First we make the variation with respect to \(\overline{\Omega}\), obtaining:
\begin{eqnarray}
    \partial_\mu\, \partial^\mu \Omega + j\gamma\, \partial_t \Omega + m^2 \Omega =0;
\label{12}
\end{eqnarray}
when the variation is made with respect to \(\Omega\), the conjugate of the Eq.(\ref{12}) is obtained. Thus, we have the equation of motion for the entire system,
\begin{small}
\begin{eqnarray}
\begin{aligned}
J^{+} \left[\partial_{\mu} \partial^{\mu}\Omega^{+} +\gamma \partial_t \Omega^{+} + m^2\Omega^{+} \right]
+ J^{-}\left [ \partial_\mu \partial^{\mu}\Omega^{-} - \gamma \partial_t \Omega^{-} + m^2\Omega^{-} \right ] =0.  \label{13}
\end{aligned}
\end{eqnarray}
\end{small}
We have explicitly obtained the corresponding part for the subsystem of interest that is accompanied by the idempotent \(J^{+}\), and the part of the environment identified with \(J^{-}\); the last one has a negative sign in the dissipative term \(\gamma\), indicating the time inversion, besides that it is the mirror copy of the system of interest \cite{33}. In \cite{25} the authors constructed a solution, which contains real exponentials, then they analytically extend that solution in the hyperbolic complex plane; it can be also be complexified to the standard complex plane, and in both cases the relative sign between \(\omega^{2}_{k}\) and \(k^{2}\) is not altered in the dispersion relation; however, there is a sign change in the dissipative term \(\gamma^{2}\), being negative (\(-\gamma^{2}\)) for the standard scheme and positive (\(+\gamma^{2}\)) for the scheme hyperbolic. Here we have the two imaginary units in this scheme, the product \(ij\) will appear in our solution and as we will see, the sign for the dissipative term will be negative as in the standard scheme. Thus, the formal solution for the Eq.(\ref{12}) has the form,
\begin{eqnarray}
    \eta(\boldsymbol{x},t)=ae^{i p_{1}x^{\mu}} e^{jp_{2}x^{\nu}} + be^{-ip_{1}x^{\mu}} e^{-jp_{2}x^{\nu}}, 
\label{14}
\end{eqnarray}
where \(a,b\) are arbitrary coefficients, and \(\omega_{1,2}\) and \(\boldsymbol{k}_{1,2}\) are real parameters. Whit this solution we extends \(\omega \rightarrow \omega_{1} + i \omega_{2}\) and \(k \rightarrow k_{1} + i k_{2}\); however, we can always find, through a Lorentz rotation, a system where we have \(k_{2}=0\), that is, the spatial imaginary part for \(k\) vanishes; hence we are only concerned with the problem of temporal dissipation. Using the property \(e^{j \chi}=e^{\chi}J^{+} + e^{-\chi}J^{-}\), and the mentioned criterion, we have that (\ref{14}) can be rewritten as \footnote{This solution can also be obtained by proposing a real exponential, \(\Omega(\vec{x},t)=e^{\omega t - \vec{k}\cdot \vec{x}}\), and promoting \((\omega , \vec{k}) \in \mathbb{H}\), and considering restrictions for obtaining convergent solutions.},
\begin{eqnarray}
    \eta(\boldsymbol{x}, t)= a \, e^{\left[(\alpha +i\omega_{1})t - i\boldsymbol{k} \cdot \boldsymbol{x}\right]} J^{+} + b \, e^{\left[(\alpha - i\omega_{1}t) + i\boldsymbol{k} \cdot \boldsymbol{x} \right]} J^{-}.
\label{15}
\end{eqnarray}
With this, we obtain the solutions for the equation of motion (\ref{13}),
\begin{eqnarray}
\begin{aligned}
\Omega^{+}=\texttt{a}_{1}\,e^{\left[\left(\Gamma_{1} +i \omega_{1} \right) t-i \boldsymbol{k}_{1} \cdot \boldsymbol{x}\right]}+ \texttt{b}_{1}\, e^{\left[\left(\Gamma_{1} -i \omega_{1}\right) t+i \boldsymbol{k}_{1} \cdot \boldsymbol{x}\right]}, \\
\Omega^{-}=\texttt{a}_{2}\,e^{\left[\left(\Gamma_{2}+i \omega_{2}\right) t-i \boldsymbol{k}_{2} \cdot \boldsymbol{x}\right]}+ \texttt{b}_{2}\, e^{\left[\left(\Gamma_{2}-i \omega_{2} \right) t+i \boldsymbol{k}_{2} \cdot \boldsymbol{x}\right]};  
\label{16}
\end{aligned}
\end{eqnarray}
where (\(\texttt{a}_{1,2}, \texttt{b}_{1,2}\)) are arbitrary coefficients and the spectral parameters (\(\omega, \boldsymbol{k}\)) are real-valued. We will first calculate the solution for the part that corresponds to the system of interest, that is, the solution corresponding to \(J^{+}\) in (\ref{13}). Especifically we can to determine of \(J^{\pm}\) projections of the Eq.(\ref{13}) and to use the property (\ref{6}); taking the solution \(\Omega^{+}\) of (\ref{16}) and performing the substitution into the equation of motion we obtain,
\begin{eqnarray}
\begin{aligned}
    &e^{[(\Gamma_{1} + i \omega_{1})t - i\boldsymbol{k}_{1}\cdot \boldsymbol{x}]} \left[ \Gamma_{1}^{2} - \omega_{1}^{2} + k_{1}^{2} + \gamma \Gamma_{1} + m^2 + i \omega_{1}( 2 \Gamma_{1}   + \gamma) \right]\\
    &+ e^{[(\Gamma_{1} - i \omega_{1})t + i\boldsymbol{k}_{1}\cdot \boldsymbol{x}]} \left[\Gamma_{1}^{2} - \omega_{1}^{2} + k_{1}^{2} + \gamma \Gamma_{1} +m ^{2} - i \omega_{1}(2 \Gamma_{1}  + \gamma)\right] =0.
\label{17}
\end{aligned}
\end{eqnarray}
A similar procedure is done for the solution \(\Omega^{-}\). From the imaginary part of these expressions, we obtain the dissipative coefficients,
\begin{equation}
\begin{aligned}
     \Gamma_{1}=-\frac{\gamma}{2}, \quad \quad \Gamma_{2}=\frac{\gamma}{2};
\label{18}     
\end{aligned}
\end{equation}
and from the real part, we obtain the dispersion relations,
\begin{eqnarray}
    \omega_{1,2}=\pm \sqrt{k_{1,2}^{2} + \underbrace{m^{2}  -\frac{\gamma^{2}}{4}}_{modified \;\; mass}}.
\label{19}
\end{eqnarray}
It is necessary to notice some aspects of the frequencies of the system; the frequencies are real \((m^2 - \frac{\gamma^2}{4}) \geq 0\) and therefore there is not an IR cut-off, for the case when \((m^2 - \frac{\gamma^2}{4}) < 0\), we take the positive values of the radicand\footnote{Otherwise, the radicand could take negative values and possibly lead to imaginary frequencies, which would imply superluminal speeds. In this work we only considered the real frequencies} in (\ref{19}), therefore \(k^2 \geq -(m^2 - \frac{\gamma^2}{4})\), obtaining an IR cut-off. Therefore, the general solution is represented as plane waves damped by a decaying factor \(e^{-\frac{\gamma}{2}t}\) for the system of interest and a growing factor \(e^{\frac{\gamma}{2}t}\) for the environment; this is in accordance with the rules of TFD, since in this context, the environment evolves in the reverse direction of time. Furthermore, we can write an arbitrary combination of solutions for the field \(\Omega\) in terms of the (\(J^{+},J^{-}\)) basis, 
\begin{eqnarray}
    \Omega(\boldsymbol{x},t)=J^{+} e^{-\frac{\gamma}{2} t}\left[\texttt{a}_{1} e^{i\left(\omega_{1}-\boldsymbol{k}_{1} \cdot \boldsymbol{x}\right)}+ \overline{\texttt{a}}_{2} e^{-i\left(\omega_{1}-\boldsymbol{k}_{1} \cdot \boldsymbol{x}\right)}\right]+J^{-} e^{\frac{\gamma}{2} t}\left[ \overline{\texttt{b}}_{1} e^{i\left(\omega_{2}-\boldsymbol{k}_{2} \cdot \boldsymbol{x}\right)}+ \texttt{b}_{2} e^{-i\left(\omega_{2}-\boldsymbol{k}_{2} \cdot \boldsymbol{x}\right)}\right], 
\label{20}
\end{eqnarray}
where \(\texttt{a}_{1,2}, \texttt{b}_{1,2}\) are hypercomplex arbitrary coefficients; this combination will allow to construct the quantum fields in the next section.

\section{Quantum fields and field commutators}
\label{sec4}
\begin{figure}[H]
    \centering
    \includegraphics[width=8.5cm, height=4.5cm]{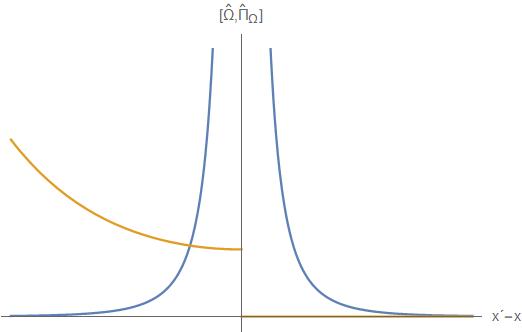}
    \caption{\footnotesize{Graphics corresponding to the real part \(\left(\frac{\sqrt{m^2 - \frac{\gamma^2}{4}}}{(x' - x)} K_{1}\right)\) of (\ref{34}); where \(K\) is a function of \((x^{\prime}-x)\) for a fixed \(m_{mod}\). The blue line corresponds to the real part and the orange line corresponds to the imaginary part. Both lines are no continuous at
\((x'-x)=0\). Note that this field commutator behaves similarly to a Dirac delta function as would be expected in standard QFT. The most notable feature between the expression (\ref{34}) and a Dirac delta is that the value of the Dirac delta is zero everywhere except at the origin, where it is infinite; in Eq.(\ref{34}) we have that at the origin it has an infinite value and it has a finite value for \((x^{\prime}-x) \neq 0\); furthermore, the commutator tends to zero as \((x^{\prime} -x) \rightarrow \infty\).}}
    \label{fig1}
\end{figure}

In the spectral decomposition used in \cite{25} the authors considered to split the range of spectral parameters \(\boldsymbol{k}\) into two parts, \((-\infty,0) \leftrightarrow J^{-}\) for the environment, and \((0, +\infty) \leftrightarrow J^{+}\) for the system of interest; this is done to avoid the divergences that arise in the field operator commutators. In contrast to this, in the formulation at hand we can take the full range \(\boldsymbol{k} \in (-\infty,+\infty)\) since, due to the presence of the additional complex unit \(i\), such divergences do not appear; therefore, the spectral decomposition for the field operator built with the solution (\ref{20}) reads,
\begin{eqnarray}
\begin{aligned}
\widehat{\Omega}(\boldsymbol{x},t)= \left\{ e^{-\frac{\gamma}{2} t} J^{+}  \right.& \int_{-\infty}^{\infty}  \left[\hat{\texttt{a}}_{1}\left(\boldsymbol{k}_{1}\right) e^{i\left(\omega_{k_{1}} t-\boldsymbol{k}_{1} \cdot \boldsymbol{x}\right)}+\hat{\texttt{a}}_{2}^{\dagger}\left(\boldsymbol{k}_{1}\right) e^{-i\left(\omega_{k_{1}} t-\boldsymbol{k}_{1} \cdot \boldsymbol{x}\right)}\right] d \boldsymbol{k}_{1} \\
&\left.\hspace{-0.9cm} + \, e^{\frac{\gamma}{2} t} J^{-} \int_{-\infty}^{\infty}  \left[\hat{\texttt{b}}_{1}^{\dagger}\left(\boldsymbol{k}_{2}\right) e^{i\left(\omega_{k_{2}} t-\boldsymbol{k}_{2} \cdot \boldsymbol{x}\right)}+\hat{\texttt{b}}_{2}\left(\boldsymbol{k}_{2}\right)  e^{-i\left(\omega_{k_{2}} t-\boldsymbol{k}_{2} \cdot \boldsymbol{x}\right)}\right] d \boldsymbol{k}_{2}\right\}. \label{21}
\end{aligned}
\end{eqnarray}
With this expression we can build the field commutator,
\begin{eqnarray}
\begin{aligned}
\left[\hat{\Omega}(\boldsymbol{x},t), \hat{\Omega}^{\dagger}\left(\boldsymbol{x}^{\prime}, t\right)\right]
&= \int_{-\infty}^{\infty} d \boldsymbol{k} \int_{-\infty}^{\infty} d \boldsymbol{k}^{\prime}\left\{J^{+}\left(e^{i\left[\left(\omega_{k}-\omega_{k'}\right) t-\boldsymbol{k} \cdot \boldsymbol{x}+\boldsymbol{k}^{\prime} \cdot \boldsymbol{x}^{\prime}\right]}\left[\hat{\texttt{a}}_{1}\left(\boldsymbol{k}\right), \hat{\texttt{b}}_{1}\left(\boldsymbol{k'}\right)\right]\right. \right.\\
&\left. \left. \left. \hspace{4cm}+e^{i\left[\left(\omega_{k}+\omega_{k'}\right) t\,-\,\boldsymbol{k} \cdot \boldsymbol{x}\,-\,\boldsymbol{k}' \cdot \boldsymbol{x}^{\prime}\right]}\left[\hat{\texttt{a}}_{1}\left(\boldsymbol{k}\right), \hat{\texttt{b}}_{2}^{\dagger}\left(\boldsymbol{k}'\right)\right]\right.\right.\right.\\
&\left.\left.\left. \hspace{3.8cm}+e^{-i\left[(\omega_{k} +\omega_{k'}) t\,-\,\boldsymbol{k} \cdot \boldsymbol{x}\,-\,\boldsymbol{k}' \cdot \boldsymbol{x}'\right]}\left[\hat{\texttt{a}}_{2}^{\dagger}\left(\boldsymbol{k}\right), \hat{\texttt{b}}_{1}\left(\boldsymbol{k}'\right)\right]\right.\right.\right.\\
&\left.\left.\hspace{4.06cm}+e^{i\left[(\omega_{k'}-\omega_{k})t\,+\,\boldsymbol{k} \cdot \boldsymbol{x}\,-\,\boldsymbol{k}' \cdot \boldsymbol{x}' \right]}  \left[\hat{\texttt{a}}_{2}^{\dagger}\left(\boldsymbol{k}\right), \hat{\texttt{b}}_{2}^{\dagger}\left(\boldsymbol{k}'\right)\right]\right)\right.\\
&\left.\hspace{2.5cm}+J^{-}\left( e^{-i\left[\left(\omega_{k}-\omega_{k'}\right) t\,+\, \boldsymbol{k}' \cdot \boldsymbol{x}\,-\, \boldsymbol{k} \cdot \boldsymbol{x}^{\prime}\right]}    \left[\hat{\texttt{b}}_{1}^{\dagger}(\boldsymbol{k}'), \hat{\texttt{a}}_{1}^{\dagger}\left(\boldsymbol{k}\right)\right] \right.\right.\\
&\hspace{3.1cm}+
e^{-i\left[\left(\omega_{k'}+\omega_{k}\right)t\,-\, \boldsymbol{k}' \cdot \boldsymbol{x}\,-\, \boldsymbol{k} \cdot \boldsymbol{x}^{\prime}\right]} \left[\hat{\texttt{b}}_{2}\left(\boldsymbol{k}'\right), \hat{\texttt{a}}_{1}^{\dagger}\left(\boldsymbol{k}\right)\right]\\
&\hspace{3.35cm}+ e^{i\left[\left(\omega_{k'}+\omega_{k}\right) t\,-\, \boldsymbol{k}' \cdot \boldsymbol{x}\,-\, \boldsymbol{k} \cdot \boldsymbol{x}^{\prime}\right]} \left[\hat{\texttt{b}}_{1}^{\dagger}\left(\boldsymbol{k}'\right), \hat{\texttt{a}}_{2}\left(\boldsymbol{k}\right)\right] \\
&\left.\left.\hspace{3.15cm}+ e^{-i\left[\left(\omega_{k'}-\omega_{k}\right) t\, -\, \boldsymbol{k}' \cdot \boldsymbol{x}\,+\,\boldsymbol{k} \cdot \boldsymbol{x}'\right]}\left[\hat{\texttt{b}}_{2}\left(\boldsymbol{k}'\right), \hat{\texttt{a}}_{2}\left(\boldsymbol{k}\right)\right] \right)\right\}.   
\label{22}
\end{aligned}
\end{eqnarray}
One of the major difficulties that appear in the study of dissipative systems in quantum mechanics that involve the duplication of fields, is that the commutation canonical relations are not preserved under the temporal evolution \cite{39}. In the commutator (\ref{22}), this time dependence has been eliminated due to the terms that have the damping factors vanish, \(e^{ -\frac{\gamma}{2}t} (J^{+} J^{ -})\left[ \hat{\texttt{a}}, \hat{\texttt{a}}^{\dagger} \right]=e^{+\frac{ \gamma}{2}t} ( J^{+} J^{-})\left[ \hat{\texttt{b}}, \hat{\texttt{b}}^{\dagger} \right] =0\), due to the property (\ref{6}). On the other hand, there is also another elimination of the damping factors, since they have the form \(e^{-\frac{\gamma}{2}}J^{+} \cdot e^{+\frac{\gamma }{2}}J^{+}\). The rest of the field commutators have a similar structure and the same characteristics as the expression (\ref{22}), thus they do not have dissipative factors; however, there exists special commutator with dissipative factors, namely, 
\begin{eqnarray}
\begin{aligned}
\left[\hat{\Omega}^{\dagger}(\boldsymbol{x},t), \hat{\Pi}_{\Omega}(\boldsymbol{x}',t)\right]&= \int_{-\infty}^{\infty} d \boldsymbol{k} \int_{-\infty}^{\infty} d \boldsymbol{k}^{\prime}\left\{e^{\gamma t}\, J^{+}\left(e^{i\left[\left(\omega_{k'}-\omega_{k}\right) t\,+\,\boldsymbol{k} \cdot \boldsymbol{x}-\boldsymbol{k}^{\prime} \cdot \boldsymbol{x}^{\prime}\right]}\left[\hat{\texttt{b}}_{1}\left(\boldsymbol{k}\right), \hat{\texttt{b}}_{2}^{\dagger}\left(\boldsymbol{k'}\right)\right]\right. \right.\\
&\left.\left.\hspace{3.8cm}-e^{-i\left[(\omega_{k'}-\omega_{k})t\,+\,\boldsymbol{k} \cdot \boldsymbol{x}\,-\,\boldsymbol{k}' \cdot \boldsymbol{x}' \right]}  \left[\hat{\texttt{b}}_{2}^{\dagger}\left(\boldsymbol{k}\right), \hat{\texttt{b}}_{1}\left(\boldsymbol{k}'\right)\right]\right)\right.\\
&\left.\hspace{2.6cm}+ e^{-\gamma t} J^{-}\left( e^{i\left[\left(\omega_{k'}-\omega_{k}\right) t\,-\, \boldsymbol{k}' \cdot \boldsymbol{x'}\,+\, \boldsymbol{k} \cdot \boldsymbol{x}\right]}    \left[\hat{\texttt{a}}_{1}^{\dagger}(\boldsymbol{k}), \hat{\texttt{a}}_{2}\left(\boldsymbol{k'}\right)\right] \right.\right.\\
&\left.\left.\hspace{3.8cm}+ e^{-i\left[\left(\omega_{k'}-\omega_{k}\right) t\, -\, \boldsymbol{k}' \cdot \boldsymbol{x'}\,+\,\boldsymbol{k} \cdot \boldsymbol{x}\right]}\left[\hat{\texttt{a}}_{2}\left(\boldsymbol{k}\right), \hat{\texttt{a}}_{1}^{\dagger}\left(\boldsymbol{k'}\right)\right] \right)\right\}. 
\label{23}
\end{aligned}
\end{eqnarray}
The expression (\ref{23}) contains the trivial canonical commutation relations \([\hat{a}^{\dagger}_{\textbf{p}}, \hat{a}^{\dagger}_{\textbf{q}}]=[\hat{a}_{\textbf{p}},\hat{a}_{\textbf{q}}]=0\), and the nontrivial relations \([\hat{a}_{\textbf{p}}, \hat{a}^{\dagger}_{\textbf{q}}]\) and \([\hat{b}_{\textbf{p}}, \hat{b}^{\dagger}_{\textbf{q}}]\); which do not vanish in general. However, in this work we will consider that the last commutators are vanishing \([\hat{a}_{\textbf{p}},\hat{a}^{\dagger}_{\textbf{q}}]=[\hat{b}_{\textbf{p}},\hat{b}^{\dagger}_{\textbf{q}}]=0\). The main reason is that, if we keep these commutators switched on, the damping coefficients \(e^{\pm \gamma t}\) do not disappear, implying divergences, which bring us back to the main problem in of the dissipative dynamics. On the other hand, assuming for a moment that we do not have divergence problems when one uses these canonical relations, the use of these commutators does not allow us to identify the entanglement suffered by the subsystems due to dissipation, since they are only describing the subsystems separately, throught the pure commutators for type-a bosons for the subsystem of interest, and pure commutators for type-b bosons for the environment. Thus, we have a very special feature of the formulation at hand, since it is essential to have the non-vanishing commutator \([\hat{\texttt{a}}(\boldsymbol{k}),\hat{\texttt{b}}(\boldsymbol{k}')]\neq 0\) in order to do not trivialize the theory.\\
\noindent
From (\ref{22}) and (\ref{23}) we need to propose the following commutation rules for the annihilation and creation operators,
\begin{eqnarray}
\begin{matrix}
\begin{array}{l}
\left[\hat{\texttt{a}}_{1,2}\left(\boldsymbol{k}\right), \hat{\texttt{b}}_{1,2}\left(\boldsymbol{k}'\right)\right]=\rho_{i} \, \delta\left(\boldsymbol{k}-\boldsymbol{k}'\right) \\
\left[\hat{\texttt{b}}_{1,2}^{\dagger}\left(\boldsymbol{k}'\right), \hat{\texttt{a}}_{1,2}^{\dagger}\left(\boldsymbol{k}\right)\right]=\overline{\rho}_{i}\, \delta\left(\boldsymbol{k}-\boldsymbol{k}'\right) \\
\left[\hat{\texttt{a}}_{1,2}\left(\boldsymbol{k}\right), \hat{\texttt{b}}_{1,2}^{\dagger}\left(\boldsymbol{k}'\right)\right]= \sigma_{j} \, \delta({\boldsymbol{k}+\boldsymbol{k}'}) \\
\left[\hat{\texttt{b}}_{1,2}\left(\boldsymbol{k}'\right), \hat{\texttt{a}}_{1,2}^{\dagger}\left(\boldsymbol{k}\right)\right]= \overline{\sigma}_{j} \, \delta(\boldsymbol{k} + \boldsymbol{k}'),\\
\left[\hat{\texttt{b}}_{1,2}\left(\boldsymbol{k}\right), \hat{\texttt{b}}_{1,2}^{\dagger}\left(\boldsymbol{k}'\right)\right]=0,\\
\left[\hat{\texttt{a}}_{1,2}\left(\boldsymbol{k}\right), \hat{\texttt{a}}_{1,2}^{\dagger}\left(\boldsymbol{k}'\right)\right]=0;
\end{array}, & \quad i,j=1,2,3,4;  \quad \quad
\begin{matrix}
[a_{1},b_{1}]\rightarrow \rho_{1},\\ 
[a_{1},b_{2}] \rightarrow \rho_{2},\\ 
[a_{2},b_{1}]\rightarrow \rho_{3},\\ 
[a_{2},b_{2}]\rightarrow \rho_{4},\\
\label{24}
\end{matrix} & 
\end{matrix}
\end{eqnarray}
where \(\rho_{i}\) and \(\sigma_{j}\) are arbitrary elements in \(\mathbb{H}\) and in general depend on \((\boldsymbol{k}, \boldsymbol{k}')\); the column on the right hand side indicates the identification (the correspondence) between the indices for the commutators and the coefficients \(\rho\). Now considering that,
\begin{eqnarray}
\left[\hat{\Omega}(\boldsymbol{x},t), \hat{\Omega}^{\dagger}(\boldsymbol{x}',t) \right]= \left[\hat{\Omega}(\boldsymbol{x},t), \hat{\Omega}^{\dagger}(\boldsymbol{x}',t) \right]^{\dagger}_{(\boldsymbol{x} \leftrightarrow  \boldsymbol{x}')} \hspace{0.5cm} \Rightarrow \quad \sigma_{j}=0;
\label{25}
\end{eqnarray}
that is, two commutation rules in (\ref{24}) vanish. Therefore, using the commutation relations (\ref{25}), the field commutator (\ref{22}) reduces to, 
\begin{eqnarray}
\begin{aligned}
\left[\widehat{\Omega}(\boldsymbol{x}, t), \widehat{\Omega}^{\dagger}\left(\boldsymbol{x}^{\prime}, t\right)\right] &= (2 \pi)^{n} \left[J^{+}(\rho_{1}- \overline{\rho}_{4})+ J^{-}(\overline{\rho}_{1}- \rho_{4})\right] \delta_n(\boldsymbol{x}'-\boldsymbol{x}).
\label{26}
\end{aligned}
\end{eqnarray}
In \cite{25} this field commutator diverges and convergence criteria are considered, by spliting the full spectral parameter \((-\infty, +\infty)\) into two parts, as mentioned at the beginning of this section. Here the integration converges due to we have the additional complex unit \(i\), which allows us to obtain the Dirac delta; this same field commutator in \cite{25} has a more complicated mathematical expression; however, in both formulations, this commutator has the same physically acceptable asymptotic limits,
\begin{eqnarray}
\begin{aligned}
      \lim_{\boldsymbol{x}'- \boldsymbol{x} \rightarrow 0}  \left[\widehat{\Omega}(\boldsymbol{x},t), \widehat{\Omega}
      ^{\dagger}(\boldsymbol{x}',t)\right] &= \infty ,&  \quad
      \lim_{\boldsymbol{x}'- \boldsymbol{x} \rightarrow \infty}  \left[\widehat{\Omega}(\boldsymbol{x},t), \widehat{\Omega}
      ^{\dagger}(\boldsymbol{x}', t)\right] &= 0.
\label{27}
\end{aligned}
\end{eqnarray}
Note that the field commutator (\ref{26}) does not depend on either the modified mass or the dissipative parameter \(\gamma\).
Now we obtain the conjugate canonical moment from Eq.(\ref{7}),
\begin{eqnarray}
\Pi_{\Omega} \equiv \frac{\partial \mathcal{L}}{\partial \dot{\Omega}}=\dot{\overline{\Omega}}-j \frac{\gamma}{2} \overline{\Omega};
\label{28}
\end{eqnarray}
then we construct the conjugate momentum operator,
\begin{small}
\begin{eqnarray}
\begin{aligned}
\widehat{\Pi}_{\Omega}(\boldsymbol{x}, t)&=- \left\{e^{\frac{\gamma}{2} t} J^{+} \int_{-\infty}^{\infty} i \omega_{k_{2}}  \left[\hat{\texttt{b}}_{1}\left(\boldsymbol{k}_{2}\right) e^{-i\left(\omega_{k_{2}} t-\boldsymbol{k}_{2} \cdot \boldsymbol{x}\right)}-\hat{\texttt{b}}_{2}^{\dagger}\left(\boldsymbol{k}_{2}\right) e^{i\left(\omega_{k_{2}} t-\boldsymbol{k}_{2} \cdot \boldsymbol{x}\right)}\right] d \boldsymbol{k}_{2}\right.\\
&\left.+ \, e^{-\frac{\gamma}{2} t} J^{-} \int_{-\infty}^{\infty} i \omega_{k_{1}}  \left[\hat{\texttt{a}}_{1}^{\dagger}\left(\boldsymbol{k}_{1}\right) e^{-i\left(\omega_{k_{1}} t-k_{1} \cdot x\right)}-\hat{\texttt{a}}_{2}\left(\boldsymbol{k}_{1}\right) e^{i\left(\omega_{k_{1}} t-k_{1} \cdot x\right)}\right] d \boldsymbol{k}_{1}\right\},
\label{29}
\end{aligned}
\end{eqnarray}
\end{small}
therefore, we have the following equal time field commutator:
\begin{eqnarray}
\begin{aligned}
\left[\widehat{\Pi}_{\Omega}(\boldsymbol{x}, t), \widehat{\Pi}_{\Omega}^{\dagger}\left(\boldsymbol{x}^{\prime}, t\right)\right]&=
 (2 \pi)^{n} \left[J^{+}(\rho_{1}- \overline{\rho}_{4})+ J^{-}(\overline{\rho}_{1}- \rho_{4})\right] \left[ {\delta}_n''\left ( \boldsymbol{{x}'}-\boldsymbol{x} \right ) - \left(m^{2} - \frac{\gamma^{2}}{4} \right) \delta_n(\boldsymbol{x}' - \boldsymbol{x})\right],  
\label{30}
\end{aligned}
\end{eqnarray}
where \(\delta''\) is the second derivative of the delta function. Again the dependence on dissipative factors \(e^{\pm \frac{\gamma}{2}t}\) has been eliminated due to the condition (\ref{6}). In contrast to \cite{25}, here we have the presence of the modified mass; in particular we have the following limits,
\begin{eqnarray}
\lim_{(m^2 - \frac{\gamma^2}{4}) \rightarrow \, 0} \left[\widehat{\Pi}_{\Omega}(\boldsymbol{x}, t), \widehat{\Pi}_{\Omega}^{\dagger}\left(\boldsymbol{x}^{\prime}, t\right)\right]= (2 \pi)^{n} \left[J^{+}(\rho_{1}- \overline{\rho}_{4})+ J^{-}(\overline{\rho}_{1}- \rho_{4})\right] {\delta}_n''\left ( \boldsymbol{{x}'}-\boldsymbol{x} \right ).
\label{31}
\end{eqnarray}
On the other hand, we have the  following limit,
\begin{eqnarray}
\begin{small}
\begin{aligned}
\lim_{m \rightarrow \, 0} \left[\widehat{\Pi}_{\Omega}(\boldsymbol{x}, t), \widehat{\Pi}_{\Omega}^{\dagger}\left(\boldsymbol{x}^{\prime}, t\right)\right]&= (2 \pi)^{n} \left[J^{+}(\rho_{1}- \overline{\rho}_{4})+ J^{-}(\overline{\rho}_{1}- \rho_{4})\right] \left({\delta}_n''\left ( \boldsymbol{{x}'}-\boldsymbol{x} \right) + \frac{\gamma^2}{4} \delta_n(\boldsymbol{x}' - \boldsymbol{x}) \right).
\label{32}
\end{aligned}
\end{small}
\end{eqnarray}
In both formulations, the one developed in \cite{25} and in the present work, the field commutator (\ref{30}) satisfies the same limits that we mentioned in equation (\ref{27}).
Similarly, we construct another equal time commutator,
\begin{small}
\begin{eqnarray}
\begin{aligned}
\left[\widehat{\Omega}(\boldsymbol{x}, t), \widehat{\Pi}_{\Omega}\left(\boldsymbol{x}^{\prime}, t\right)\right]&=
-i \int_{-\infty}^{\infty}  \omega_{\boldsymbol{k}}\left[\left(J^{+}\rho_{1}+ J^{-}\bar{\rho}_{1}\right) e^{i \boldsymbol{k} \cdot\left(\boldsymbol{x}^{\prime}-\boldsymbol{x}\right)}+\left(J^{+}\bar{\rho}_{4}-J^{-}\rho_{4}\right) e^{-i \boldsymbol{k} \cdot\left(\boldsymbol{x}^{\prime}-\boldsymbol{x}\right)}\right] d \boldsymbol{k}.
\label{33}
\end{aligned}
\end{eqnarray}
\end{small}
\noindent
This field commutator is the only one (of the three ones in this formulation) that does not vanish in both, the standard scheme and in the present scheme. In the standard scheme this field commutator is simply a Dirac delta, without providing more information. The integral in the commutator (\ref{33}) can be solved both numerically and analytically, considering certain restrictions for each case. We solve this integral in one spatial dimension (or \textit{1+1} background space-time), and due to the presence of the frequencies, we use the dispersion relation (\ref{19}) with \((m^2 - \frac{\gamma^2}{4}) > 0\), for obtaining,
\begin{eqnarray}
\left[\widehat{\Omega}(x, t), \widehat{\Pi}_{\Omega}\left(x^{\prime}, t\right)\right]= - 2 i \left[J^{+}(\rho_{1}- \overline{\rho}_{4})+ J^{-}(\overline{\rho}_{1}- \rho_{4})\right] \frac{\sqrt{m^2 -\frac{\gamma^2}{4}}}{({x}' - {x})} K_{1} \left(\left | m^2 -\frac{\gamma^2}{4} \right | ({x}' - {x}) \right),
\label{34}
\end{eqnarray}
where \(K_{1}(x)\) is the modified Bessel function. The integral (\ref{33}) does not converge in general. It can also be solved numerically when the constraint \((m^2 - \frac{\gamma^2}{4})<0\) is considered. On the other hand, we can see in the Fig.\ref{fig1} the behavior of the commutator (\ref{34}), for the the following cases: it converges to zero as \(m_{mod} \rightarrow +\infty\) and the field commutator diverges as \(m_{mod} \rightarrow 0\).
\noindent
Similar to the limit computed for the field commutator (\ref{31}), we also compute the limit when \((m^2 - \frac{\gamma^2}{4}) \rightarrow 0\) for the field commutator (\ref{34}), which shows us that this limit diverges. On the other hand, we can leave the \(\gamma\)-parameter intact and see the following limits,
\begin{eqnarray}
    \lim_{m \rightarrow 0} \left[\widehat{\Omega}(x, t), \widehat{\Pi}_{\Omega}\left(x^{\prime}, t\right)\right] = \left[J^{+}(\rho_{1}- \overline{\rho}_{4})+ J^{-}(\overline{\rho}_{1}- \rho_{4})\right] |\gamma| K_{1}\left(\frac{\gamma^4}{16} (x'-x), \right)
\label{35}
\end{eqnarray}
where we have the explicit dependence of the dissipative parameter \(\gamma\). We also have the following limits,
\begin{eqnarray}
\begin{aligned}
\lim_{m \rightarrow \infty} \left[\widehat{\Omega}(x, t), \widehat{\Pi}_{\Omega}\left(x^{\prime}, t\right)\right] &= 0, \quad \quad   &  \lim_{(m^2 - \frac{\gamma^2}{4}) \rightarrow \infty} \left[\widehat{\Omega}(x, t), \widehat{\Pi}_{\Omega}\left(x^{\prime}, t\right)\right] &=0.
\label{36}
\end{aligned}
\end{eqnarray}
Again, in both formulations, namely in \cite{25} and in this work, the field commutator (\ref{34}) satisfies the limits (\ref{27}). 
\begin{figure}[H]
    \centering
    \includegraphics[width=7.5cm, height=4.5cm]{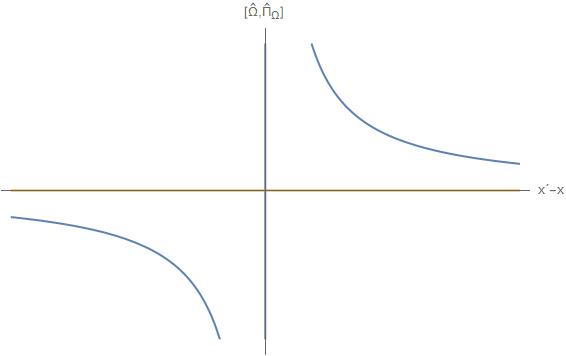}
    \caption{\footnotesize{Graphics corresponding to the commutator \(\left[\widehat{\Omega}, \widehat{\Pi}_{\Omega}\right]\) in (\ref{34}) for a fixed \((x'-x) > 0\). The orange line corresponds to the imaginary part and the blue line represents the real part; both lines are continuous throughout the real interval. For this field commutator, in the work \cite{25} the imaginary part vanishes in this region \(x'-x > 0\), and as commented, the imaginary part that does not vanish and it will lead to a commutator belonging to the extended ring.}} 
    \label{fig2}
\end{figure}
\noindent
We can see that the integrals defining all field commutators depend in general on the spatial dimension; moreover, the field commutators (\ref{30}) and (\ref{33}) have an explicit dependency on \(\gamma\) through the modified mass and the frecuencies \(\omega_{k}\); however, the field commutator (\ref{26}) does not depend on any of these quantities.

\section{ Evolution operator and the charge operator}
\label{sec5}
In \cite{25}, the function \(G(k,k^{\prime};system)\) is introduced, and depends on the spatial configurations for the subsystem and the environment; this function is real, which leads to maintain a double integration. In the formulation at hand we obtain a similar function, also identified with the total geometry of the system; our function now contains the imaginary unit \(i\), which allows us to simplify the Hamiltonian operator, as discussed in \cite{25}; we will describe this simplification for specific geometries. We start with the classical Hamiltonian operator,
\begin{eqnarray}
    H= \int\left[ 2 \Pi_{\Omega} \Pi_{\bar{\Omega}}+ \frac{1}{2} \partial_{i} \Omega \partial_{i} \bar{\Omega}+j \frac{\gamma }{2}\left(\bar{\Omega} \Pi_{\bar{\Omega}}-\Omega \Pi_{\Omega}\right) + \left(m^2 -\frac{\gamma^{2}}{4}\right) \frac{\Omega \bar{\Omega}}{2}\right]dx^{d};  
\label{37}
\end{eqnarray}
each term is a \(U(1)\times SO(1,1)\)-invariant; and again, the dissipative factors \(e^{\pm \gamma t}\) do not appear due to property (\ref{6}). Thus the Hamiltonian operator is,
\begin{small}
\begin{eqnarray}
\begin{aligned}
\hat{\mathcal{H}}\left(\gamma; t \right)= \int_{-\infty}^{\infty} d \boldsymbol{k}  \int_{-\infty}^{\infty} d \boldsymbol{k}' \left\{H_{\gamma} \, G(\boldsymbol{k}, \boldsymbol{k}';\gamma;t) \left[J^{+} \left\{\hat{\texttt{a}}_{1} (\boldsymbol{k}), \hat{\texttt{b}}_{1} (\boldsymbol{k}')\right\} + J^{-} \left\{\hat{\texttt{b}}_{2} (\boldsymbol{k}'), \hat{\texttt{a}}_{2} (\boldsymbol{k})\right\} \right] + h.c \right\},
\label{38}
\end{aligned}
\end{eqnarray}
\end{small}
where the following complex functions have been defined,
\begin{eqnarray}
\begin{aligned}
G(\boldsymbol{k},\boldsymbol{k}';\gamma;t) & \equiv e^{i (\omega_{k} - \omega_{k'})t} \int_{\underset{system}{complete}} e^{-i \boldsymbol{x} \cdot (\boldsymbol{k} - \boldsymbol{k}')} d\boldsymbol{x}^{n}=e^{i (\omega_{k} - \omega_{k'})t} \, \texttt{I}\left(\boldsymbol{k - k'} \right),
\label{39}
\end{aligned}
\end{eqnarray}
\begin{eqnarray}
\begin{aligned}
H_{\gamma} & \equiv  2\, \omega_{k'} \omega_{k} + \frac{1}{2} \boldsymbol{k'} \cdot \boldsymbol{k} + \frac{i \gamma}{2} (\omega_{k'} +\omega_{k})+ \frac{1}{2} \left(m^{2} - \frac{\gamma^{2}}{4}\right).
\label{40}
\end{aligned}
\end{eqnarray}
In (\ref{38}) appears the conjugate of (\ref{39}) and (\ref{40}). The function \(\texttt{I}(\boldsymbol{k}-\boldsymbol{k}')\) and its conjugate, are defined as integrals over the total system, and they will depende on the geometry.

\subsection{The Hamiltonian in \((1+1)\) space-time}
\label{sec5.1}
\begin{figure}[H]
\centering
\includegraphics[width=11.5cm, height=5.5cm]{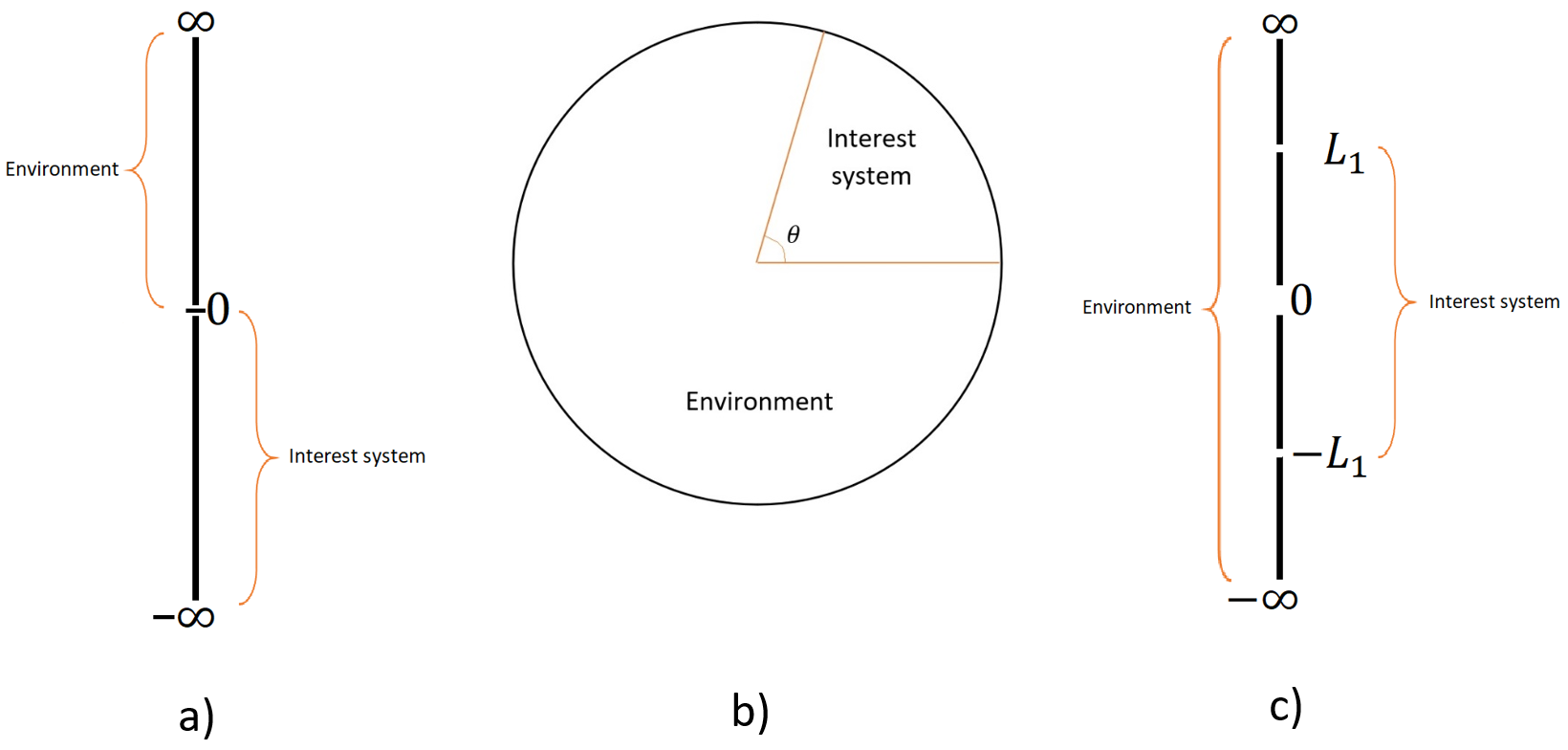}
\caption{Some examples of one-dimension geometries  for the total system.}
\label{fig3}
\end{figure}

In the expression (\ref{38}) there is not yet a defined geometry for the total system. Different geometries can be considered; some specific configurations are shown in Fig.\ref{fig3} (see \cite{24.1} for other geometries). We focus first on the a)-geometry in the Fig.\ref{fig3}.
\noindent
Therefore, the integration interval for the momenta (\ref{37}) will be \((-\infty, \infty)\). Hence, the \(\texttt{I}(\boldsymbol{k}-\boldsymbol{k}')\) functions will be reduced to Dirac deltas; thus the Hamiltonian operator takes the form,
\begin{eqnarray}
\begin{aligned}
\hat{\mathcal{H}}\left(m;\gamma;t \right)= \int_{-\infty}^{\infty} d \boldsymbol{k} \left\{ H_{\gamma}  \left[J^{+} \left\{\hat{\texttt{a}}_{1} \left(\boldsymbol{k}\right), \hat{\texttt{b}}_{1}\left(\boldsymbol{k} \right) \right\} + J^{-} \left\{\hat{\texttt{b}}_{2}\left(\boldsymbol{k}\right), \hat{\texttt{a}}_{2}\left(\boldsymbol{k}\right)\right\}\right]+ h.c \right\}.
\label{Ham41}
\end{aligned}
\end{eqnarray}
Due to the presence of the complex unit \(i\) the Hamiltonian operator could be reduced to a single \(k\)-integration, as we mentioned at the beginning of this section; however, we see that the Dirac delta only arises when we have an infinite total system. Later we will consider finite geometries, where the finite geometry of the total system will play an important role.\\

\noindent
The unitary evolution operator can be constructed as the exponential of the Hamiltonian operator by using the complex unit \(i\) or, as in the case of \cite{25}, with the complex unit \(j\). We can also think of constructing the evolution operator with a hypercomplex exponential, that is, containing the two complex units \(e^{ij \hat{\mathcal{H}}}\); however, we will use the exponential with the standard unit \(i\), since the hyperbolic complex unit in the exponential is always implicit due to the bases \(J^{+,-}\). Thus our time evolution operator is,
\begin{eqnarray}
\begin{aligned}
e^{i\hat{\cal{H}}t} & \equiv
e^{i t \int d\boldsymbol{k} \left[ H_{{\gamma}^{+}} \left ( J^{+} \{\hat{\texttt{a}}_1(\boldsymbol{k}),\hat{\texttt{b}}_1(\boldsymbol{k})\} + J^{-} \{\hat{\texttt{b}}_2(\boldsymbol{k}),\hat{\texttt{a}}_2(\boldsymbol{k})\}\right) + H_{\gamma^{-}} \left ( J^{+} \{\hat{\texttt{b}}^{\dagger}_{2}(\boldsymbol{k}),\hat{\texttt{a}}^{\dagger}_{2}(\boldsymbol{k})\} + J^{-} \{\hat{\texttt{a}}_{1}^{\dagger}(\boldsymbol{k}),\hat{\texttt{b}}_{1}^{\dagger}(\boldsymbol{k})\} \right )\right]}\\
&=J^{+} e^{it \int d\boldsymbol{k} \left[H_{\gamma^{+}} \{\hat{\texttt{a}}_1(\boldsymbol{k}),\hat{\texttt{b}}_1(\boldsymbol{k})\} \, + \, H_{\gamma^{-}} \{\hat{\texttt{b}}^{\dagger}_2(\boldsymbol{k}),\hat{\texttt{a}}^{\dagger}_2(\boldsymbol{k})\}\right]} + J^{-} e^{it \int d\boldsymbol{k}\left[ H_{\gamma{+}} \{\hat{\texttt{b}}_{2}(\boldsymbol{k}),\hat{\texttt{a}}_{2}(\boldsymbol{k})\}  \, + \, H_{\gamma^{-}} \{\hat{\texttt{a}}_{1}^{\dagger}(\boldsymbol{k}),\hat{\texttt{b}}_{1}^{\dagger}(\boldsymbol{k})\} \right]},
\label{42}
 \end{aligned}
\end{eqnarray}
where the decomposition property \(e^{j \chi}=e^{\chi}J^{+} + e^{-\chi}J^{-}\), and the annihilation property (\ref{6}), have been used. We can see that with this operator, the transition to the statistical case can be made.

\subsection{The charge operator} 
\label{section 5.2}
The following conservation law follows from the invariance of the Lagrangian under the action of \(U(1) \times SO(1,1)\), 
\begin{eqnarray}
\partial_{t}(\bar{\Omega} \dot{\Omega}-\Omega \dot{\overline{\Omega}}+j \gamma \Omega \overline{\Omega})+\partial_{i}\left(\overline{\Omega} \partial^{i} \Omega-\Omega \partial^{i} \overline{\Omega}\right)=0;
\label{43}
\end{eqnarray}
with the charge density \(j_{0}=(\overline{\Omega} \dot{\Omega}-\Omega \dot{\overline{\Omega}}+j \gamma \Omega \overline{\Omega})\) and the current \( j^{i}=\left(\bar{\Omega} \partial^{i} \Omega-\Omega \partial^{i} \overline{\Omega}\right)\).
Note the term \(j \gamma \Omega \overline{\Omega}\) in the charge density, which is a new term that appears due to dissipation. Therefore, the charge \(Q\) associated with this hypercomplex current is given by the following space integral,
\begin{eqnarray}
    Q=\int j_{0}\, dx^{n}=\int\left(\overline{\Omega} \Pi_{\overline{\Omega}}-\Omega \Pi_{\Omega}\right) dx^{n};
\label{44}
\end{eqnarray}
in terms of the four real components of the fields, the charge is:
\begin{eqnarray}
    Q= \int \left[ i (\dot{\phi}_{2} \phi_{1} - \dot{\phi}_{1}\phi_{2} + \dot{\psi}_{1} \psi_{2} - \dot{\psi}_{2} \psi_{1}) + j (\dot{\psi}_{1} \phi_{1} + \dot{\psi}_{2} \phi_{2} - \dot{\phi}_{1} \psi_{1} - \dot{\phi}_{2} \psi_{2}) \right] \, dx^{n},
\label{45}
\end{eqnarray}
note that in the limit when \(\psi_{1}=\psi_{2}=0\), the \(U(1)\)-charge is recovered, which is given by \(i(\dot{\phi_{2}}\phi_{1} - \dot{\phi_{1}}\phi_{2})\). Using the expressions (\ref{21}) and (\ref{28}) of the fields in terms of the \(\hat{\texttt{a}}, \hat{\texttt{a}}^{\dagger}, \hat{\texttt{b}}, \hat{\texttt{b}}^{\dagger}\) operators, the charge operator can be written as,
\begin{eqnarray}
\begin{aligned}
    \hat{Q}\left(\gamma ; t\right)&=-j \int \left[\left\{\widehat{\Omega}, \widehat{\Pi}_{\Omega}\right\}j +c . c\right] d x^{n}\\
&=-2 i  \int_{-\infty}^{\infty}  \omega_{k} \left[ J^{+} \left\{\hat{\texttt{a}}_{1}\left(\boldsymbol{k}\right), \hat{\texttt{b}}_{1}\left(\boldsymbol{k}\right)\right\} + J^{-} \left\{\hat{\texttt{a}}_{1}^{\dagger}\left(\boldsymbol{k}\right), \hat{\texttt{b}}_{1}^{\dagger}\left(\boldsymbol{k}\right)\right\} \right.  \\
&\left. \hspace{2.85cm}-\left( J^{+} \left\{\hat{\texttt{b}}_{2}^{\dagger} \left(\boldsymbol{k}\right), \hat{\texttt{a}}_{2}^{\dagger} \left(\boldsymbol{k}\right)\right\} +  J^{-}\left\{\hat{\texttt{b}}_{2} \left(\boldsymbol{k}\right), \hat{\texttt{a}}_{2} \left(\boldsymbol{k}\right)\right\} \right) \right] d \boldsymbol{k}.
\label{46}
\end{aligned}
\end{eqnarray}
We can observe that the charge operator is anti-Hermitian \(\hat{Q}^{\dagger}=-\hat{Q}\). We remark the differences with respect to the results in \cite{25}. First notice that in our expression (\ref{45}) we have a \(\boldsymbol{k}\)-integral, since, once again, the imaginary unit \(i\) allows us to perform an integration; on the other hand, in our charge operator there are two additional anticommutators \(\left\{\hat{\texttt{b}}_{2}^{\dagger}, \hat{\texttt{a}}_{2}^{\dagger}\right\}\), and \(\left\{\hat{\texttt{b}}_{2}, \hat{\texttt{a}}_{2}\right\}\), which correspond to the copy-system.

\section{The vacuum in \(\mathbb{H}\)}
\label{section 6}
According to \cite{25} and \cite{18} a definition with linear expressions in operators of annihilation, leads to a trivial QFT. The vacuum will be defined as a coherent state for the following operators that involve two annihilation operators (the derivation of this definition of vacuum is extensive and can be seen in detail in \cite{18}), 
\begin{eqnarray}
\begin{aligned}
J^{+}\left \{ \hat{\texttt{a}}_{1}(\boldsymbol{k}),\hat{\texttt{b}}_{1}(\boldsymbol{k}') \right \}\ket{0}&=
J^{+}\lambda_1 \ket{0}, \quad  & J^{-}\left \{ \hat{\texttt{b}}_2(\boldsymbol{k}'),\hat{\texttt{a}}_2(\boldsymbol{k}) \right \}\ket{0}&=J^{-}\lambda_2 \ket{0}
\label{47}
\end{aligned}
\end{eqnarray}
for all \(\boldsymbol{k}, \boldsymbol{k}'\) and \(\lambda_{1,2}\) are elements of the ring \(\mathbb{H}\). With subscript 1 we are representing the definition of the vacuum for the subsystem of interest and with subscript 2 for the environment. Now, considering the quadratic combination of the creation and annihilation operators in the observables (\ref{Ham41}) and (\ref{46}) on the vacuum state, we have,
\begin{small}
\begin{eqnarray}
\begin{aligned}
&J^{-}\left \{ \hat{\texttt{a}}_1(\boldsymbol{k}),\hat{\texttt{b}}_1(\boldsymbol{k}') \right \}^{\dagger} \ket{0}
= J^{-} \hat{\texttt{a}}^{\dagger}_{1}(\boldsymbol{k}) \hat{\texttt{b}}^{\dagger}_{1}(\boldsymbol{k}') \ket{0} + J^{-}\hat{\texttt{b}}^{\dagger}_{1}(\boldsymbol{k}') \hat{\texttt{a}}^{\dagger}_{1}(\boldsymbol{k}) \ket{0} = J^{-}\left(\ket{^{1}\texttt{a}_{k}, ^{1}\texttt{b}_{{k}'}} + \ket{^{1}\texttt{b}_{{k}'},^{1}\texttt{a}_{{k}}} \right),\\
&J^{+}\left \{ \hat{\texttt{b}}_2(\boldsymbol{k}'),\hat{\texttt{a}}_2(\boldsymbol{k}) \right \}^{\dagger} \ket{0}= J^{+}\left(\ket{^{2}\texttt{b}_{{k}'},^{2}\texttt{a}_{k}} + \ket{^{2}\texttt{a}_{k'},^{2}\texttt{b}_{k}} \right).
\label{48}
\end{aligned}    
\end{eqnarray}
\end{small} 
To obtain the vacuum expectation values corresponding to the Hamiltonian and the charge operator we use the expression (\ref{47}),
\begin{eqnarray}
\begin{aligned}
\braket{0| \hat{\cal{H}}|0}&= \braket{0|0} \int_{-\infty}^{\infty} d\boldsymbol{k} H_{\gamma} \left[  \left( J^{+}\lambda_1 + c.c.\right) + \left( J^{-}\lambda_2  + c.c.\right) \right],
\label{49}
\end{aligned}
\end{eqnarray}
\begin{eqnarray}
\begin{aligned}
\braket{0|\hat{Q}|0}= - & \braket{0|0} \int_{\infty}^{\infty} d\boldsymbol{k}\left[ \left( J^{+} \lambda_{1} + c.c.\right) - \left(J^{-}\lambda_2 +c.c. \right)\right].
\label{50}
\end{aligned}
\end{eqnarray}
The integrals in (\ref{49}) and (\ref{50}) are divergent, but these divergences can be eliminated. To visualize this, we write the part on the right hand side of (\ref{47}) as,  
\begin{align}
    J^{+}\lambda_1 +c.c.&=\lambda_{\texttt{I}} + \lambda_{\texttt{II}}, &
    J^{-}\lambda_2 + c.c.&= \lambda_{\texttt{III}} - \lambda_{\texttt{IV}}.
\label{51}
\end{align}
Now we can write to (\ref{49}) and (\ref{50}) in terms of \(\lambda_{\texttt{I,II,III,IV}}\), and separating the real part and imaginary, we have,
\begin{eqnarray}
\begin{aligned}
\braket{0| \hat{\cal{H}}|0}&=  \braket{0|0} \int_{-\infty}^{\infty} d\boldsymbol{k} \left[H_{k'}  \left(\lambda_{\texttt{I}} + \lambda_{\texttt{II}} +\lambda_{\texttt{III}} - \lambda_{\texttt{IV}} \right) + 2ij\omega_{k}\gamma \left(\lambda_{\texttt{I}} + \lambda_{\texttt{II}} - \lambda_{\texttt{III}} + \lambda_{\texttt{IV}}\right) \right],
\label{52}
\end{aligned}
\end{eqnarray}
\begin{eqnarray}
\begin{aligned}
\braket{0|\hat{Q}|0}= -2i & \braket{0|0} \int_{\infty}^{\infty} d\boldsymbol{k} \left( \omega_{k}\right)\left[\lambda_{\texttt{I}} + \lambda_{\texttt{II}} - \lambda_{\texttt{III}} + \lambda_{\texttt{IV}}\right];
\label{53}
\end{aligned}
\end{eqnarray}
where \(H_{k'}\) is,
\begin{eqnarray}
    H_{k'}=2\omega_{k}^{2} + \frac{1}{2}k^{2} + \frac{1}{2} \left(m^{2}-\frac{\gamma^{2}}{4} \right).
\label{54}
\end{eqnarray}
Therefore, by imposing the restrictions,
\begin{eqnarray}
  \lambda_{\texttt{I}} + \lambda_{\texttt{II}} =0, & \hspace{1cm}
  \lambda_{\texttt{III}}-\lambda_{\texttt{IV}}=0;
\label{lambdaZero}
\end{eqnarray}
 it means, imposing the vanishing of the projections (\ref{51}), hence, the vanishing of the v.e.v'.s (\ref{52}) and (\ref{53}) are achieved. The first condition corresponds to the case of \(\lambda_{1} \sim J^{-}\) and the second corresponds to \(\lambda_{2} \sim J^{+}\) in the definitions (\ref{47}), where \(\sim\) means proportional to.

\section{Entangled states}
\label{section 7}
Considering the action of the evolution operator in the form (\ref{42}) on the vacuum state defines above, one obtains, 
\begin{eqnarray}
\begin{aligned}
\ket{0(t)} &\equiv e^{i\hat{\mathcal{H}}t} \ket{0}\\ &=e^{i\int_{-\infty}^{\infty} d\boldsymbol{k}|\left[H_{\gamma^{+}} \left(J^{+} \left\{\hat{\texttt{a}}_{1}(\boldsymbol{k}), \hat{\texttt{b}}_{1}(\boldsymbol{k})\right\} + J^{-} \left\{\hat{\texttt{b}}_{2}(\boldsymbol{k}), \hat{\texttt{a}}_{2}(\boldsymbol{k})\right\}\right) + H_{\gamma^{-}} \left( J^{+} \left\{\hat{\texttt{b}}_{2}^{\dagger}(\boldsymbol{k}), \hat{\texttt{a}}_{2}^{\dagger}(\boldsymbol{k})\right\} + J^{-} \left\{\hat{\texttt{a}}_{1}^{\dagger}(\boldsymbol{k}), \hat{\texttt{b}}_{1}^{\dagger}(\boldsymbol{k})\right\} \right) \right] t} \ket{0};
\label{56}
\end{aligned}
\end{eqnarray}
if the original vacuum is normalized, the evolved state (\ref{56}) will also be a normalized state for any \(t\):
\begin{eqnarray}
\braket{0(t)|0(t)} \equiv \bra{0} e^{-i \hat{\mathcal{H}}t} \cdot e^{i \hat{\mathcal{H}}t} \ket{0}= \braket{0|0}.
\label{57}
\end{eqnarray}
As well known in the quantum mechanics the representations of the canonical commutation relations are all unitarily equivalent to each other (Stone-Von Neuman theorem); thus the evolved vacuum leaves the original space of states, in a finite volume case; since \(t \rightarrow \infty\), this yields an asymptotic state orthogonal to the initial state \(\ket{0}\) \cite{42}. On the other hand, in a QFT the number of degrees of freedom is infinite, thus, there are infinite unitarily non-equivalent representations of the canonical commutation relations, this allows describing different systems that can be in different phases \cite{43}. Using this, a dissipative quantum model of the brain has been studied where there is a non-unit time evolution \cite{44}. Considering the evolved vacuum (\ref{56}), we see that this state evolves always aligned with the original vacuum state, 
\begin{eqnarray}
\begin{aligned}
\braket{0|0(t)}&= e^{\int_{-\infty}^{\infty} d\boldsymbol{k} \left[H_{\gamma^{+}} J^{+} \lambda_{1} - c.c. \right]t - \int_{-\infty}^{\infty} d\boldsymbol{k} \left[H_{\gamma^{+}} J^{-} \lambda_{2} - c.c. \right]t} \braket{0|0}\\
&\small{=\braket{0|0} \exp\left\{\int_{-\infty}^{\infty} d\boldsymbol{k} \left[ iH_{k} \left(\lambda_{\texttt{I}} + \lambda_{\texttt{II}} + \lambda_{\texttt{III}} - \lambda_{\texttt{IV}} \right) + j2\omega_{k} \gamma \left(\lambda_{\texttt{I}} + \lambda_{\texttt{II}} + \lambda_{\texttt{III}} - \lambda_{\texttt{IV}} \right) \right] t \right\}} \neq 0.
\label{58}
\end{aligned}
\end{eqnarray}
The usual imaginary part and the hyperbolic imaginary part of the second line in Eq.(\ref{58}) have been separated. The expression in (\ref{58}) has the form of a bi-complex phase \(e^{i \alpha} e^{j\beta}\), then the property (\ref{4}) can be used, obtaining
\begin{eqnarray}
\begin{aligned}
\braket{0|0(t)}&=\braket{0|0} \left(\cos[\alpha(t)] \cosh[\beta(t)] + i \sin[\alpha(t)] \cosh[\beta(t)] + j \cos[\alpha(t)] \sinh[\beta(t)] + ij \sin[\alpha(t)] \sinh[\beta(t)]\right),
\label{59}
\end{aligned}
\end{eqnarray}
where,
\begin{eqnarray}
\begin{aligned}
\beta(t)&= 2\gamma t \int_{-\infty}^{\infty} d\boldsymbol{k} \, \omega_{k} \left(\lambda_{\texttt{I}} + \lambda_{\texttt{II}} + \lambda_{\texttt{III}} - \lambda_{\texttt{IV}}\right)=0,\\
\alpha(t)&= t\int_{-\infty}^{\infty} d\boldsymbol{k} \, H_{k} \left(\lambda_{\texttt{I}} + \lambda_{\texttt{II}} + \lambda_{\texttt{III}} - \lambda_{\texttt{IV}} \right)=0,
\label{60}
\end{aligned}
\end{eqnarray}
the vanishing is due to the condition (\ref{lambdaZero}), which ensure that the temporal evolution does not leave the original Hilbert space. Furthermore, as we know, the proportionality conditions \((\lambda_{1} \sim J^{-}), (\lambda_{2} \sim J^{+})\), will lead to that the entanglement dynamics is constructed in both directions, namely, for the subsystem of interest \((J^{+})\) and, for the environment \((J^{-})\) perspective. With these constraints in mind, and using the expansion \(e^{J^{\pm} \chi}=1 + J^{\pm} \sum_{n=1}^{\infty} \frac{\chi^{n}}{n!}\), we have that the evolved state (\ref{56}) can be rewritten as,
\begin{eqnarray}
\begin{aligned}
\ket{0(t)}&= \exp{ \left\{ t\int_{\infty}^{\infty} d\boldsymbol{k} \, H_{\gamma} \left(J^{+} \left\{\hat{\texttt{b}}_{2}^{\dagger}, \hat{\texttt{a}}_{2}^{\dagger}\right\} - J^{-} \left\{\hat{\texttt{a}}_{1}^{\dagger}, \hat{\texttt{b}}_{1}^{\dagger}\right\}       \right)  \right\}} \ket{0}\\
&= \ket{0} + J^{+} t \int_{-\infty}^{\infty} d \boldsymbol{k} \, H_{\gamma} \left(\ket{^{2}\texttt{b}_{k}, ^{2}\texttt{a}_{{k}}} + \ket{^{2}\texttt{a}_{{k}},^{2}\texttt{b}_{{k}}} \right) - J^{-} t \int_{-\infty}^{\infty} d \boldsymbol{k} \, H_{\gamma}\left(\ket{^{1}\texttt{a}_{k}, ^{1}\texttt{b}_{{k}}} + \ket{^{1}\texttt{b}_{{k}},^{1}\texttt{a}_{{k}}} \right) + ...;
\label{61}
\end{aligned}
\end{eqnarray}
the notation (\ref{47}) for excited states is used. This state is then entangled in moments, since it cannot be factorized into the product of single modes. In the next section we will see that entanglement is present even in the absence of dissipation.

\subsection{Entangled asymptotic states }
\label{section7.1}
In this section, we return to the general Hamiltonian operator (\ref{38}), in which the geometry of the total system has been not specified; now, we consider different geometrical configurations and we shall construct the entangled states associated. We begin by considering the following evolved state,
\begin{eqnarray}
\begin{aligned}
\ket{0(t)} &\equiv e^{i\hat{\mathcal{H}}t} \ket{0} =\exp\left\{{i{\int d\boldsymbol{k}}^{+\infty}_{-\infty} {\int d\boldsymbol{k}'}\left[ H_{\gamma} G \left( J^{+} \left\{\hat{\texttt{b}}_{2}^{\dagger}(\boldsymbol{k}'), \hat{\texttt{a}}_{2}^{\dagger}(\boldsymbol{k})\right\} + J^{-} \left\{\hat{\texttt{a}}_{1}^{\dagger}(\boldsymbol{k}), \hat{\texttt{b}}_{1}^{\dagger}(\boldsymbol{k}')\right\} \right) \right] t}\right\} \ket{0},
\label{62}
\end{aligned}
\end{eqnarray}
where we used \(\mathcal{\hat{H}}\) given in (\ref{38}), and the function \(G\) is defined in (\ref{39}). Also, we have used the conditions given in Eq.(\ref{lambdaZero}), that is, the eigenvalues in Eq.(\ref{47}) have disappeared. The above expression can be rewritten as,
\begin{eqnarray}
\begin{aligned}
e^{i\hat{\mathcal{H}}t} \ket{0} =\exp\left\{i{\int d\boldsymbol{k}}^{+\infty}_{-\infty} {\int d\boldsymbol{k}'}\; H_{\gamma}(\boldsymbol{k},\boldsymbol{k}')\,t \,e^{i(\omega_{k'}-\omega_{k})t} \, \texttt{I}^{+}(\boldsymbol{k}-\boldsymbol{k}') \, \hat{\texttt{W}}(\boldsymbol{k},\boldsymbol{k}') \right\} \ket{0},
\label{63}
\end{aligned}
\end{eqnarray}
where,
\begin{eqnarray}
\hat{\texttt{W}}(\boldsymbol{k},\boldsymbol{k}') \equiv J^{+} \left\{\hat{\texttt{b}}_{2}^{\dagger}(\boldsymbol{k}'), \hat{\texttt{a}}_{2}^{\dagger}(\boldsymbol{k})\right\} + J^{-} \left\{\hat{\texttt{a}}_{1}^{\dagger}(\boldsymbol{k}), \hat{\texttt{b}}_{1}^{\dagger}(\boldsymbol{k}')\right\},
\label{64}
\end{eqnarray}
and \(\texttt{I}(\boldsymbol{k}-\boldsymbol{k}')\) is given in Eq.(\ref{39}) and is depending also on the geometry of the total system. Since we have incorporated the imaginary unit \(i\) in the present work, this has a great implication, since it opens the way for us to begin to study the ergodicity of systems that undergo entanglement, a characteristic that contrasts with \cite{25}, where it is not possible to explore this area. As it is well known, asymptotic states are closely related to the ergodic theory; this theory has various applications in different areas of mathematics \cite{50}\cite{51} and physics \cite{52}\cite{53}, where ergodicity criteria and the physical properties are established for specific configurations of a system. These configurations can have well-defined asymptotic states (states that reach thermal equilibrium) or cyclostationary asymptotic states (states that do not thermalize, but relax to periodic states) \cite{45}. This section is the beginning of a long way to go to study the ergodicity in entangled quantum systems with a new approach, the hypercomplex formalism. An important aspect to consider in the expression (\ref{63}) is that, when integrating over the total system, the expression \(\texttt{I}(\boldsymbol{k}-\boldsymbol{k}'\)) will have different forms when considering different geometries. An example, if the total system has the geometry of the area of a disk, then the integration will give us a modified Bessel function, which must be considered when performing the \(\boldsymbol{k}\)-integrations in (\ref{63}).

\subsection{An asymptotic entangled state for a finite total system}
\label{section7.2}
\begin{figure}[H]
    \centering
    \includegraphics[width=3.5cm, height=4.0cm]{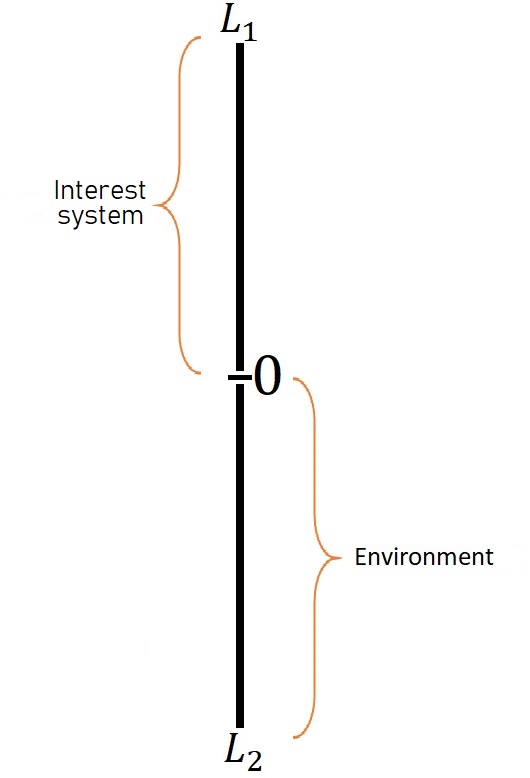}
    \caption{Total system contained in a finite interval.}
    \label{fig4}
\end{figure}

For this subsection we consider the total one-dimensional system as represented in Fig.\ref{fig4} with finite ranges, that is, the range for the environment is \((L_{2}, 0)\) and the range for the system of interest is \((0, L_{1})\), thus the total system is contained in the interval \((L_{1}, L_{2})\).
\noindent
For the state (\ref{62}), a distribution for a discrete time sequence can be identified, namely \(t e^{i(\omega_{k'} -\omega_{k})t}\), as the generating function of the delta function \(\delta_{t}[i(\omega_{k'}-\omega_{k})]\); this delta sequence satisfies:
\begin{eqnarray}
\lim_{n \rightarrow \infty} \int_{-\infty}^{\infty} \delta_{n}(is) f(s) ds = -if(0).
\label{65}
\end{eqnarray}
Hence, we have the following asymptotic state with dissipation,
\begin{eqnarray}
\begin{aligned}
\lim_{t \rightarrow \infty} e^{i \hat{\mathcal{H}}t} \ket{0}&= \exp\left\{i \lim_{t \rightarrow \infty}{\int dk}^{+\infty}_{-\infty} {\int dk'}\; H_{\gamma}(k,k')\,t \,e^{i(\omega_{k'}-\omega_{k})t} \, \texttt{I}^{+}(k-k') \, \hat{\texttt{W}}(k,k') \right\} \ket{0}\\
&=\exp \left\{5 \left(L_2 - L_1 \right) \int_{-\infty}^{\infty} dk \left[ \frac{\omega_{k}^{3}}{k}   - i \frac{2 \omega^{2}_{k} \gamma}{k}\right] \hat{\texttt{W}}(k,k)  \right\} \ket{0},
\label{66}
\end{aligned}
\end{eqnarray}
where the function \(H_{\gamma}(k,k')\) given in Eq.(\ref{40}) has been considered, and the function \(\texttt{I}(k-k')\) reduces, in this case, to \((L_2 -L_1)\), which is the total length of the one-dimensional system; in addition, the real part and the imaginary part have been separated. Furthemore, using the expansion \(e^{J^{\pm} \chi}=1 + J^{\pm} \sum_{n=1}^{\infty} \frac{\chi^{n}}{n!}\), and defining,
\begin{eqnarray}
\eta_{k} \equiv \frac{\omega_{k}^{3}}{k}   - i \frac{2 \omega^{2}_{k} \gamma}{k},
\label{67}
\end{eqnarray}
the state (\ref{66}) can be described as,
\begin{eqnarray}
\begin{aligned}
\lim_{t \rightarrow \infty} e^{i \hat{\mathcal{H}}t} \ket{0}&= 
\ket{0} + 5J^{+} (L_2 - L_1) \int_{-\infty}^{\infty}dk \; \eta_{k} \, \left( \ket{^{2}\texttt{b}_{k}, ^{2}\texttt{a}_{k}} + \ket{^{2}\texttt{a}_{k}, ^{2}\texttt{b}_{k}}\right) + \cdots  \\
& \hspace{1.5cm} +\, 5J^{-} (L_2 - L_1) \int_{-\infty}^{\infty}dk \; \eta_{k} \, \left( \ket{^{1}\texttt{a}_{k}, ^{1}\texttt{b}_{k}} + \ket{^{1}\texttt{b}_{k}, ^{1}\texttt{a}_{k}}\right) +\cdots
\end{aligned}
\label{68}
\end{eqnarray}
where the notation given in Eq.(\ref{48}) for excited states has been used. In expression (\ref{68}) we can identify what the subsystem of interest and the environment observe, by considering the projections of (\ref{68}) on the basis \((J^{+}, J^{-})\). For the point of view of the subsystem of interest, we project \(J^{+}\) on Eq.(\ref{68}),
\begin{eqnarray}
J^{+}\lim_{t \rightarrow \infty} e^{i \hat{\mathcal{H}}t} \ket{0}= 
J^{+} \left( \ket{0} + 5 (L_2 - L_1) \int_{-\infty}^{\infty}dk \; \eta_{k} \, \left( \ket{^{2}\texttt{b}_{k}, ^{2}\texttt{a}_{k}} + \ket{^{2}\texttt{a}_{k}, ^{2}\texttt{b}_{k}}\right) + \cdots \right);
\label{69}
\end{eqnarray}
therefore, the subsystem of interest observes creation of bosons of type \(\texttt{a}\) and type \(\texttt{b}\), corresponding to the environment. Now, if we project Eq.(\ref{68}) on the basis \(J^{-}\), we will have the point of view of the environment, where it observes creation of bosons corresponding to the subsystem of interest, which are represented with the superscript 1. Furthermore, we also note that this state is then entangled in the momenta, since it cannot be factored into the product of individual modes.
Now, if the dissipative parameter \(\gamma\) disappears, the following asymptotic state is obtained,
\begin{eqnarray}
\lim_{t \rightarrow \infty} e^{i \hat{\mathcal{H}}t} \ket{0}=\exp\left\{5 (L_{2}-L_{1}) \int_{-\infty}^{\infty} dk\, \frac{(k^{2}+m^{2})^{3/2}}{k}\hat{\texttt{W}}(k,k) \right\} \ket{0};
\label{70}
\end{eqnarray}
where \(\hat{\texttt{W}}(k, k')\) has been defined in Eq.(\ref{64}). Similarly to Eq.(\ref{68}), an expansion can be done for the state (\ref{70}), obtaining,
\begin{eqnarray}
\begin{aligned}
\lim_{t \rightarrow \infty} e^{i \hat{\mathcal{H}}t} \ket{0}&= \ket{0} + 5J^{+} (L_2 - L_1) \int_{-\infty}^{\infty}dk \; \frac{(k^2+m^2)^{3/2}}{k} \left( \ket{^{2}\texttt{b}_{k}, ^{2}\texttt{a}_{k}} + \ket{^{2}\texttt{a}_{k}, ^{2}\texttt{b}_{k}}\right) + \cdots  \\
&  +\, 5J^{-} (L_2 - L_1) \int_{-\infty}^{\infty}dk \; \frac{(k^2+m^2)^{3/2}}{k} \left( \ket{^{1}\texttt{a}_{k}, ^{1}\texttt{b}_{k}} + \ket{^{1}\texttt{b}_{k}, ^{1}\texttt{a}_{k}}\right) +\cdots; 
\label{71}
\end{aligned}
\end{eqnarray}
similarly to (\ref{69}), we can project the state (\ref{70}) on the basis \((J^{+}, J^{-})\) to obtain the point of view of the subsystem of interest and the environment, respectively. Furthermore, we can observe that the entanglement is present even in the absence of dissipation, since the large number of degrees of freedom present in free quantum field theories induce entanglement with other degrees of freedom along the boundary \cite{47}.\\


\subsection{An asymptotic entangled state for an infinite total system}
\begin{figure}[H]
    \centering
    \includegraphics[width=3.5cm, height=4.0cm]{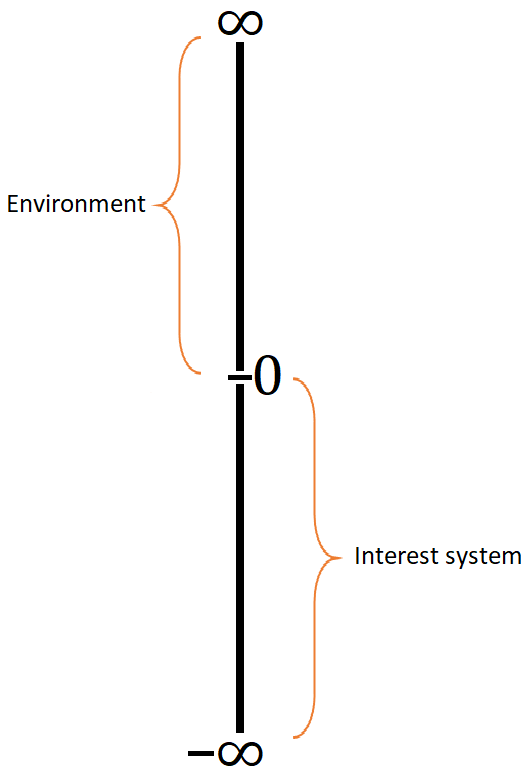}
    \caption{Total system contained in an infinite interval.}
    \label{fig5}
\end{figure}

Now we consider the geometry in which the one-dimensional subsystems are semi-infinite; this spatial configuration is represented in the Fig.\ref{fig5}.
\noindent
For this case one can show that the function \(\texttt{I}^{+}(k-k')\) reduces to a Dirac delta; thus the state in (\ref{62}) takes the form,
\begin{eqnarray}
\begin{aligned}
\lim_{t \rightarrow \infty} e^{i \hat{\mathcal{H}}t} \ket{0}&= \exp\left\{5  \pi \lim_{t \rightarrow \infty} t \int_{-\infty}^{\infty} dk \left[\frac{\gamma \omega_{k}}{5} + i \omega_{k}^{2} \right] \hat{\texttt{W}}(k,k)  \right\} \ket{0}\\
&=\exp\left\{\lim_{t \rightarrow \infty} \gamma t \int_{-\infty}^{\infty} dk \, \omega_{k} \, \hat{\texttt{W}}(k,k) \right\} \left[\cos{\left(5 \pi \lim_{t \rightarrow \infty} t \int_{-\infty}^{\infty} dk \; \omega_{k}^{2} \, \hat{\texttt{W}}(k,k) \right)} + i \sin{(...)} \right],
\label{72}
\end{aligned}
\end{eqnarray}
where the sine function has the same argument as the cosine. There is no an asymptotic state defined, since in general the expression (\ref{72}) diverges. However, there are works where these asymptotic states are analyzed in more detail \cite{33}, where the authors consider that the product \(\gamma t\) to be finite, giving a holographic interpretation between two AdS boundaries; from this, they make a comparision between their results that were applied to the entropy of the system with, the results of finite temperature, interpreting \(\gamma\) as temperature, which suggests ergodicity in the limit thermodynamic; obtaining as a result that, the holographic dual on this two asymptotically AdS boundaries of his theory, corresponds to the BTZ black hole and where the fields are definied. On the other hand, it can be seen that when \(t=0\) in the expression (\ref{72}), the usual vacuum is obtained. Furthermore, if the dissipative parameter \(\gamma\) vanishes in Eq.(\ref{72}), an oscillating state is obtained, that is, a cyclostationary state,
\begin{eqnarray}
\lim_{t \rightarrow \infty} e^{i \hat{\mathcal{H}}t} \ket{0}= \exp\left\{5 i \pi \lim_{t \rightarrow \infty} t \int_{-\infty}^{\infty} dk  \left(k^{2} + m^{2}\right) \hat{\texttt{W}}(k,k)  \right\} \ket{0},
\label{73}
\end{eqnarray}
and analogously to the expression (\ref{68}), we can project the state (\ref{72}) on the basis \((J^+, J^-)\) to have the point of view of the subsystem of interest or the environment; for the subsystem of interest, one obtains,
\begin{eqnarray}
  J^{+} \lim_{t \rightarrow \infty} e^{i \hat{\mathcal{H}}t} \ket{0}= J^{+} \exp\left[5 i \pi \lim_{t \rightarrow \infty} t \int_{-\infty}^{\infty} dk  \left(k^{2} + m^{2}\right) \left(\ket{^{2}\texttt{b}_{{k}'},^{2}\texttt{a}_{k}} + \ket{^{2}\texttt{a}_{k'},^{2}\texttt{b}_{k}} \right)  \right] \ket{0}.
\label{74}
\end{eqnarray}
We can notice that, when we have the point of view of the subsystem of interest, we see bosons corresponding to the environment (subscript 2) and when we observe from the perspective of the environment, we see bosons that correspond to the subsystem of interest (subscript 1). In \cite{33} the corresponding states with \(\gamma=0\) and \(t=0\) are equivalent; however, we see that in this work they are totally different; a more detailed explanation can be seen in \cite{25}. The oscillation occurs when the systems are semi-infinite and the fields are free. As the authors comment in \cite{45}, the absence of thermalization is unusual when dissipation is absent; explaining it with an example, with the Langevin equation for a free Brownian particle of mass \(m\), where by using the condition of zero integral friction, that is, an asymptotic condition on the dissipation kernel in the Laplace domain, they can show the absence of thermalization when \(\gamma=0\). In addition, this is accompanied by another phenomenon, the superdiffusion.

\section{The usual weighted measure}
\label{wmeasure}
Dealing with quantum field theories it is usual to see that the expansion used for the momentum contains the weight \(\frac{1}{\sqrt{\omega_{k}}}\) as measure of integration, a convenient choice for the normalization of coefficients of type \(a_{\textbf{p}}\). This choice imposes the well-known equal-time commutation relation \(\left[\phi(t, \textbf{x}), \Pi(t, \textbf{y})\right]= i \delta^{3}(\textbf{x}- \textbf{y}) \). In the development of this work, we did not impose this weight on our field; however, if we follow the usual path found in the literature, we will not obtain significant changes in our field commutators (\ref{26}), (\ref{30}) and (\ref{33}). In this section we will discuss the results obtained by adding this weight.\\

\noindent
Considering the weight \(\frac{1}{\sqrt{\omega_{k}}}\) in the measure for the field (\ref{21}) and its moment (\ref{29}), we obtain,
\begin{eqnarray}
    \left[\hat{\Omega}(\boldsymbol{x},t), \hat{\Omega}^{\dagger}(\boldsymbol{x}',t)\right]&= 2 \left[J^{+}(\rho_{1} - \overline{\rho}_{4}) + J^{-}(\overline{\rho}_{1} - \rho_{4})\right] K_{0}\left( \left|m^2 - \frac{\gamma^2}{4}\right| (\boldsymbol{x}' - \boldsymbol{x}) \right),
\label{75}
\end{eqnarray}
\begin{eqnarray}
    \left[\hat{\Pi}_{\Omega}(\boldsymbol{x},t), \hat{\Pi}^{\dagger}(\boldsymbol{x}',t)\right]&= -2 \left[J^{+}(\rho_{1} - \overline{\rho}_{4}) + J^{-}(\overline{\rho}_{1} - \rho_{4})\right] \frac{\sqrt{m^2 - \frac{\gamma^2}{4}}}{(x'-x)} K_{1}\left(\left|m^2 - \frac{\gamma^2}{4}\right| (x'-x)\right),
\label{76}
\end{eqnarray}
\begin{eqnarray}
    \left[\hat{\Omega}(\boldsymbol{x},t), \hat{\Pi}_{\Omega}(\boldsymbol{x}',t)\right]&=  i \, \delta_{n}(\boldsymbol{x}' - \boldsymbol{x})\left[J^{+}(\rho_{1} - \overline{\rho}_{4}) + J^{-}(\overline{\rho}_{1} - \rho_{4})\right].
\label{77}
\end{eqnarray}
Where \(K_{0}(x)\) and \(K_{1}(x)\) are modified Bessel functions of the second kind. We can notice that, by adding the weight \((1/\sqrt{\omega_{k}})\), the value of the field commutator \([\hat{\Omega}, \hat{\Omega}^{\dagger}]\)(Eq.(\ref{26})) reduces to the field commutator \([\hat{\Omega}, \hat{\Pi}_{\Omega}]\) above, that is \([\hat{\Omega},\hat{\Omega}^{\dagger}]\) \(\rightarrow\) \([\hat{\Omega}, \hat{\Pi}_{\Omega}]\). Similarly we have the mapping \([\hat{\Pi}_{\Omega},\hat{\Pi}^{\dagger}_{\Omega}]\) \(\rightarrow\) \([\hat{\Omega}, \hat{\Pi}_{\Omega}]\) (Eq.(\ref{33})), etc. 
\begin{figure}[H]
    \centering
    \includegraphics[width=15cm, height=4.5cm]{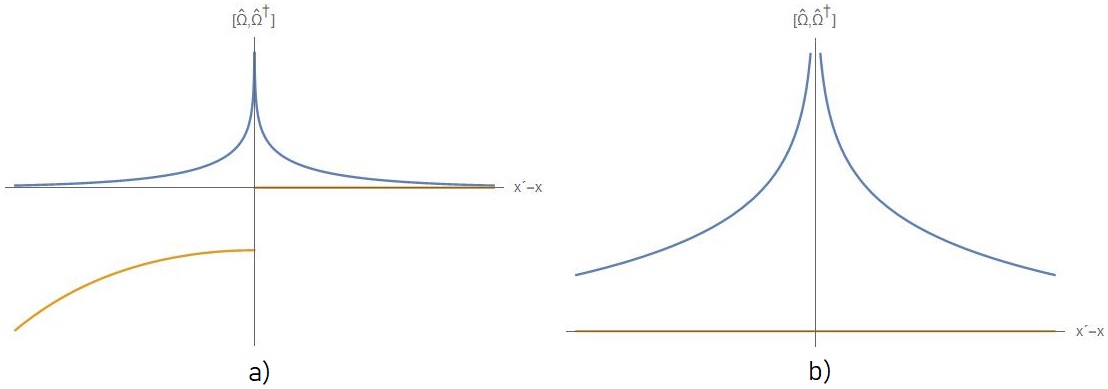}
    \caption{\footnotesize{Graph corresponding to the field commutator \([\hat{\Omega}, \hat{\Omega}^{\dagger}]\) in (\ref{74}). The blue line corresponds to the real part and the orange line corresponds to the imaginary part. In graph a), \(K\) is a function of \((x'-x)\) for a fixed \(m_{mod}\) and in graph b), \((x'-x)>0\) is fixed.}} 
    \label{fig6}
\end{figure}
\noindent
The integration process of (\ref{76}) was the same as that performed to obtain (\ref{34}), considering \((1+1)\) background space-time.
\begin{figure}[H]
    \centering
    \includegraphics[width=15cm, height=4.5cm]{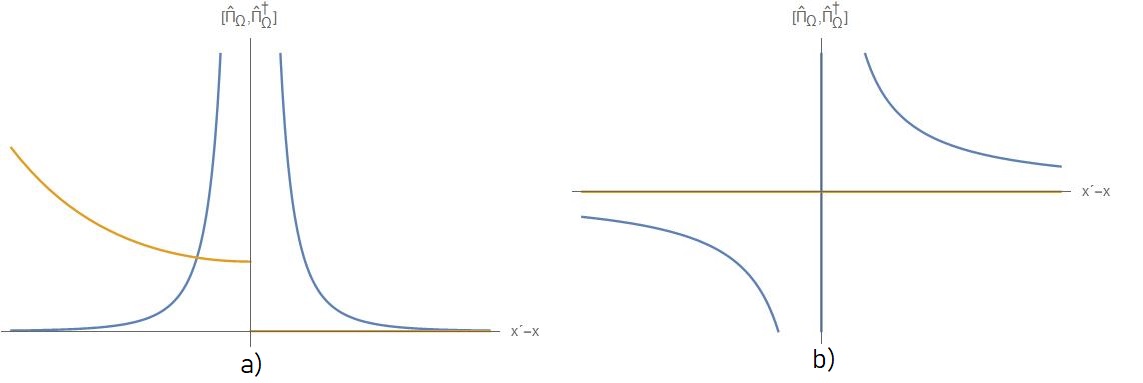}
    \caption{\footnotesize{Graph corresponding to the field commutator \([\hat{\Pi}_{\Omega}, \hat{\Pi}_{\Omega}^{\dagger}]\) in (\ref{76}). These graphs are the same as those obtained for the field commutator given in (\ref{34}). The blue line corresponds to the real part and the orange line corresponds to the imaginary part. In graph a), \(K\) is a function of \((x'-x)\) for a fixed \(m_{mod}\) and in graph b), \((x'-x)>0\) is fixed.}} 
    \label{fig6}
\end{figure}
\noindent
Thus, by using different measures, the field commutators are interchanged to each other; in the case of the field commutator \([\hat{\Omega}, \hat{\Omega}^{\dagger}]\), we get a more pronounced ``peak" for the second choice for the measure. However, this weight in the integration measure is important, since the exponent of this weight can lead to different theories, even for closed systems \cite{50}. By considering the new measure, it is enough to add the corresponding weight in quantities such as the Hamiltonian operator Eq.(\ref{38}), the charge operator Eq.(\ref{46}), etc; (moreover, the adding of the new weight does not help to eliminate the divergences in the observables Eq.(\ref{52}) and Eq.(\ref{53})) and in the entangled states.


\section{Conclusions} 
\label{conclusions}
In this work we have made an alternative formulation to study open systems whose construction is based on a hypercomplex ring, which contains two imaginary units, namely, the standard unit \(i\) and the hyperbolic unit \(j\); generalizing the formalism used in \cite{25}. The algebraic structure of the ring in the formulation in hand has led us to a non-canonical theory which, in turn, is committed to the emergence of a new \(SO(1,1)\) symmetry.
The consequences that this non-canonical theory throws at us are: the quantization scheme at hand gives us the opportunity to cure the pathologies that come from the standard quantization, since we have two complex units, the system does not leave the original Hilbert space, independly of the volume of the system. This leads to having field commutators with different characteristics from standard field commutators, since our field commutators have an explicit dependence on the dissipative parameter and the background dimension. Another implication of the formulation at hand is the way in which the grand partition function is constructed due to the emergence of the new continuous symmetry; moreover, the term corresponding to the chemical potential includes an adjustment due to dissipation (see \cite{25}); with this, we would have new descriptions of the thermal field theory using, the formalism of a hypercomplex ring; works along these research lines are in progress.\\

\noindent
On the other hand, we have shown that the vacuum state temporally evolves as an entangled state, independly of whether dissipation is on or off. This result is motivation for future works, some of which are already under development. One of these works consists of using these entangled states to calculate the time-dependent entanglement entropy, with which we can have a tool to measure the entanglement in dissipative systems. We also review the asymptotic entangled states for finite and infinite systems and, due to the structure of the Hamiltonian operator that contains the geometric information of the system, the ergodic theory naturally arises. Here we have an element that is important when studying the ergodicity of a system, the geometry. With this element we begin to introduce ourselves to the study of the ergodic behavior of these asymptotic entangled states; however, it is not enough to be able to say if a system reach thermal equilibrium or remains in a cyclo-steady state. This result is motivation to continue investigating how ergodicity is involved in our formulation; a work that we are currently developing.

\end{document}